\documentclass[aps,prc,reprint,superscriptaddress,amsmath,amssymb,floatfix,nofootinbib]{revtex4-2}

\usepackage{graphicx}
\usepackage{dcolumn}
\usepackage{bm}
\usepackage{xcolor}
\usepackage[colorlinks=true, linkcolor=gray]{hyperref}
\usepackage[normalem]{ulem} 
\usepackage{cleveref}
\usepackage[dvipsnames]{xcolor}
\usepackage{lineno}

\DeclareMathOperator{\Ree}{Re}
\DeclareMathOperator{\Imm}{Im}

\newcommand{\Kn}{\mathrm{Kn}}
\newcommand{\Rey}{\mathrm{Re}}
\newcommand{\Maa}{\mathrm{Ma}}

\begin{document}

\title{A relativistic von Kármán relation for assessing applicability of hydrodynamics}

\author{Yuuka Kanakubo}
\email{yuukakanakubo@lbl.gov}
\affiliation{RIKEN Center for Interdisciplinary Theoretical and Mathematical
Sciences (iTHEMS), RIKEN, Wako, Saitama 351-0198, Japan}
\affiliation{Nuclear Science Division, Lawrence Berkeley National Laboratory, 1 Cyclotron Road, Berkeley, CA 94720, USA}
\affiliation{Department of Physics, University of California, Berkeley, CA 94720, USA}

\author{Volker Koch}
\email{vkoch@lbl.gov}
\affiliation{Nuclear Science Division, Lawrence Berkeley National Laboratory, 1 Cyclotron Road, Berkeley, CA 94720, USA}

\date{\today}

\begin{abstract}
Relativistic viscous hydrodynamics is a widely used theoretical framework for simulating quark–gluon plasma (QGP) dynamics in relativistic heavy-ion collisions. One open question is how to quantitatively assess the validity of a hydrodynamic description of such a system. In this work, we extract the poles of linear response functions from numerical simulations of hydrodynamics and Boltzmann transport. By comparing these responses as a function of the rescaled wavenumber \(\bar{k}\), we identify the wavenumber range over which hydrodynamics captures the underlying microscopic dynamics. We show that the averaged Knudsen number, \(\langle\mathrm{Kn}^{*}\rangle\), which we define using the von Kármán relation, serves as a measure of \(\bar{k}\). We find that \(\langle\mathrm{Kn}^{*}\rangle_{\rm max}\sim 1\) marks the approximate upper limit of agreement between the linear responses of second-order viscous hydrodynamics and Boltzmann transport.
\end{abstract}

\maketitle

\section{Introduction}
Properties of a quark-gluon plasma (QGP) have been studied through 
relativistic heavy-ion collisions at the Relativistic Heavy Ion Collider (RHIC) at Brookhaven National Laboratory and the Large Hadron Collider (LHC) at CERN.
The mainstream theoretical models used for model-to-data comparisons of soft observables are so-called hybrid models, which describe the evolution of QGP with relativistic hydrodynamics and the evolution of the dilute hadron gas in the late stage with hadronic transport \cite{Heinz:2005bw,Gale:2013da}.
Pioneering works with Bayesian parameter estimation
have demonstrated that the comparison of the final hadron distribution in momentum-space can constrain the temperature dependence of the ratios of shear and bulk viscosity to entropy density, $\eta/s(T)$ and $\zeta/s(T)$, within their hybrid models \cite{Bernhard:2019bmu,JETSCAPE:2020mzn,Nijs:2020ors}.

Despite the success of the hybrid models, the inherent problem of multi-stage dynamical models remains to be resolved -- discontinuity in the dynamics among the different stages sequentially described by different theoretical descriptions.  
In the early stage of relativistic heavy-ion collisions, the problem is that energy-momentum tensors of the quarks and gluons obtained by initial state models can have a large non-equilibrium contribution while the viscous part of the energy-momentum tensor is often ignored in the initial state of conventional hydrodynamic models \cite{Oliinychenko:2015lva,Liu:2015nwa}. 
One way to resolve the problem is to describe the pre-equilibrium dynamics with QCD kinetic theory \cite{Kurkela:2018wud,Du:2025hyk}
in order to match the energy-momentum tensor between the initial gluon distribution obtained from Color Glass Condensate and the initial condition of viscous hydrodynamics. 

For the switch from hydrodynamics to hadronic transport in the later stage, Monte-Carlo sampling of hadrons with the Cooper-Frye formula \cite{PhysRevD.10.186} from a hydrodynamic hypersurface
is performed in conventional hybrid models.
The non-equilibrium part of the energy-momentum tensor in hydrodynamics is incorporated through the non-equilibrium correction of the distribution functions in the Cooper-Frye formula \cite{Pratt:2010jt,Romatschke:2003ms,Molnar:2014fva,McNelis:2019auj,McNelis:2021acu}, so that the matching of the full energy-momentum tensor is ensured.

Therefore, in both cases for the early and late stages, the possible mismatch of energy-momentum tensor can be resolved, yet the discontinuity persists in the dynamics itself.
For example, previous studies show that there is a large gap in $\eta/s$ between those at the switching temperature in hydrodynamics and those obtained from a hadronic transport \cite{Demir:2008tr,Rose:2017bjz,Ozvenchuk:2012kh,Wesp:2011yy}.
Several transport models yield $\eta/s$ values five to ten times larger than the typical values used in hydrodynamic models around the switching temperature \cite{Niemi:2015bpj,JETSCAPE:2020mzn,Nijs:2020ors}.
This may be partially circumvented by matching the $\eta/s$ at the switching temperature with the one obtained from hadronic transport model \cite{Song:2010aq} if the viscosity in hadronic transport were small enough. 
If the viscosity is so large that the system falls outside the applicability of the conventional second order viscous hydrodynamics, one needs to consider extending the applicability of viscous hydrodynamics to describe non-equilibrium dynamics as much as possible \cite{Ke:2022tqf,El:2009vj,Jaiswal:2013vta}.
Note also that the matching of viscosity means the consistency in the linear response only at the zero wavenumber limit. In general, one needs to consider the matching of linear response also at finite wavenumbers.
However, to the best of our knowledge, this problem has received relatively little attention despite the fact that quantitative extraction of viscosity with Bayesian analysis has become standardized.

In parallel to the progress in numerical studies, there has been significant progress on understanding the fundamental theoretical aspects of relativistic viscous hydrodynamics and the relation with underlying microscopic theories in analytical studies (see, $e.g.$ Refs.~\cite{Florkowski:2017olj,Schlichting:2019abc,Soloviev:2021lhs}).
The observation of QGP-like signals in small systems \cite{Schenke:2021mxx,Nagle:2018nvi} further motivated the studies 
to scrutinize the applicability of viscous hydrodynamics to many-body systems and its limitations \cite{Ambrus:2022qya,Wagner:2023jgq}. 
One way to address the applicability of hydrodynamics is the linear response theory.
The linear response theory allows us to analytically study how the system responds to a perturbation. 
Comparing the linear responses obtained by viscous hydrodynamics and Boltzmann transport theory, which are macroscopic and microscopic descriptions, reveals the range of wavenumbers over which viscous hydrodynamics can capture the underlying microscopic dynamics \cite{Romatschke:2015gic,Kurkela:2017xis,Pu:2009fj,Kamata:2020mka,Ochsenfeld:2023wxz}.
The linear response naturally depends on the viscosity or coupling strength. 
Therefore, rescaling the linear response to factor out this expected dependence
helps to identify the fundamental difference of the linear responses obtained from different theories. 
It was shown in Ref.~\cite{Du:2023bwi} that the linear response exhibits universality over different viscosities and couplings when the response functions are expressed in appropriately rescaled wavenumber and time, $i.e.,$ 
\begin{align}
\label{eq:scaled_vals}
    \bar{k} = k\eta/(sT), \quad
    \bar{t} = t (sT)/\eta,
\end{align}
where $T$ is a temperature, and $\eta/s$ is a ratio of shear viscosity to entropy density.

In this work, we demonstrate the method of investigating the linear response from hydrodynamics and transport with numerical simulations, and show that the extracted poles of the linear response show universality when the wavenumber and time are rescaled as Eq.~\eqref{eq:scaled_vals} based on Ref.~\cite{Du:2023bwi}. 
To investigate the linear response in numerical simulations, we set the initial energy density with a uniform equilibrium background and a small perturbation on a simulation grid with periodic boundary conditions.
The oscillation and damping of the perturbation contain information about the poles of the linear response functions. 
Similar analyses have been done, for instance, in Refs.~\cite{Steinheimer:2013gla,Ambru__2018,TMEP:2021ljz}.

The rescaled linear response functions can be expressed in terms of the rescaled wavenumber $\bar{k}$.
Comparisons of the rescaled linear response functions between hydrodynamics and Boltzmann transport theory identify the range of $\bar{k}$ where the hydrodynamics effectively describes Boltzmann transport theory.

In general fluid dynamics, the Reynolds and Knudsen numbers are often used as reference values when the applicability of hydrodynamics is discussed. 
We investigate the relation between the relativistic definition of the Reynolds and Knudsen numbers \cite{Denicol:2012cn,Niemi:2014wta} and the rescaled wavenumber $\bar{k}$ to see if either of them can be a good measure of $\bar{k}$.
Then we demonstrate that the averaged Knudsen number $\langle \Kn^* \rangle$ obtained from the relativistic von Kármán relation, which differs from the commonly-discussed relativistic Knudsen number, can be a good candidate which may pin down the applicability of IS hydrodynamics.
We propose \(\langle \Kn^* \rangle\) as a convenient diagnostic for assessing whether hydrodynamic simulations of relativistic heavy-ion collisions are applied within their regime of applicability. The value of \(\langle \Kn^* \rangle\) can be calculated solely from hydrodynamic variables, without knowing the wavenumber of the initial perturbation, and remains robust across the choices of time-averaging windows.

This paper is organized as follows:
In Sec.~\ref{sec:theoretical_description}, we briefly review the linear response functions from relativistic viscous hydrodynamics and the Boltzmann equation in the relaxation time approximation, which will be the analytical reference results to be compared to the results obtained from numerical simulations.
In Sec.~\ref{sec:methods}, we explain the numerical implementation of relativistic hydrodynamics and transport that are used in this paper,
and describe the methodology of extracting poles of linear responses
from numerical simulations.
In Sec.~\ref{sec:results}, we first show the results for the extracted poles of linear responses from hydrodynamics and transport in terms of the rescaled wavenumber, and compare them to the analytical reference discussed in Sec.~\ref{sec:theoretical_description}.
We then discuss the relation between several hydrodynamic dimensionless numbers and $\bar{k}$, and propose that the averaged Knudsen number $\langle \Kn^* \rangle$ obtained from the relativistic von Kármán relation, which differs from the Knudsen number commonly-used in relativistic hydrodynamics, is a good measure of $\bar{k}$. With the obtained relation between $\langle \Kn^* \rangle$ and $\bar{k}$, we quantify the upper limit of $\langle \Kn^* \rangle_{\rm max} $ at which the linear responses from hydrodynamics and transport theory start to deviate.

Throughout this paper, we use natural units, $\hbar= c = k_B = 1$, 
and the Minkowski metric, $g_{\mu\nu} = \rm diag(1, -1, -1, -1)$.

\section{Theoretical description}
\label{sec:theoretical_description}
Our goal is to study the linear response from numerical simulations.
In this section, we briefly review the linear response functions from 
relativistic viscous hydrodynamics in the Navier-Stokes and Israel-Stewart formalisms, and relativistic Boltzmann equation in the relaxation time approximation, with a 1-D perturbation in a (3+1)D system in Cartesian coordinates.

\subsection{Relativistic viscous hydrodynamics}

\subsubsection{Basic equations}
Hydrodynamics is a macroscopic theory which describes the space-time evolution of thermodynamic variables defined in the local rest frame of fluid elements (see for a review $e.g.$ \cite{Romatschke:2009im,Rocha:2023ilf}).
In this study, we investigate the linear response in a massless particle system as a first step to evaluate our methodology.
For a massless particle system, the conservation equation from relativistic hydrodynamics can be written as,  
\begin{equation}
\label{eq:hydroEq}
    \partial_\mu T^{\mu\nu}
    = \partial_\mu (
    e u^\mu u^\nu
    -
    P \Delta^{\mu\nu}
    +
    \pi^{\mu\nu}) = 0,
\end{equation}
where $P$ is pressure, $u^\mu$ is the four-velocity, $\pi^{\mu\nu}$ denotes the shear-stress tensor, and $\Delta^{\mu\nu} = g^{\mu\nu} - u^\mu u^\nu$ which is constructed to be orthogonal to $u^\mu$.
The four velocity of a fluid $u^\mu$ is chosen in the direction of the energy flux, $i.e.$, $T^\mu_\nu u^\nu = e u^\mu$, where $u^\mu$ is a normalized time-like eigenvector of the energy-momentum tensor $T^{\mu\nu}$ and $e$ is an eigenvalue which is interpreted as energy density. This particular choice of $u^\mu$ is called the Landau frame, which is typically adopted for simulations of QGP at the LHC energies.

In the first-order Navier-Stokes formalism, the shear-stress tensor is given by 
\begin{align}
\label{eq:NS}
\pi^{\mu\nu}_{\mathrm{NS}} = 2\eta \nabla^{\left\langle \mu \right.} u^{\left.\nu\right\rangle },
\end{align}
where $\eta$ is the shear viscosity, and $\nabla^\mu = \Delta^{\mu}_\nu \partial^\nu$. The traceless-symmetric projection of a tensor is denoted by the angle brackets $\langle.\rangle$ as,
\begin{align}
    A^{\langle \mu\nu \rangle} &= \Delta^{\mu\nu}_{\alpha\beta} A^{\alpha\beta} \\
    & = \left[ \frac{1}{2} (\Delta^{\mu}_\alpha \Delta^{\nu}_{\beta} + \Delta^{\nu}_\alpha \Delta^{\mu}_\beta) - \frac{1}{3}\Delta^{\mu\nu}\Delta_{\alpha\beta} \right] A^{\alpha\beta}.
\end{align}
This simple Navier-Stokes equation is known to violate causality which causes a related instability. 
The Israel-Stewart (IS) equations are second-order viscous hydrodynamic equations, which respect causality by introducing the relaxation time of dissipative currents.
The dynamics of the shear stress tensor from the simplified IS formalism \cite{Israel:1976tn,Israel:1979wp}
is given by:
\begin{align}
\label{eq:IS}
    \tau_\pi \dot{\pi}^{\mu\nu} = \pi^{\mu\nu}_{\mathrm{NS}} - \pi^{\mu\nu},
\end{align}
where the relaxation time is given by $\tau_\pi = 5\eta/(4P)$ obtained from the Boltzmann equations within RTA for a massless gas \cite{Romatschke:2011qp,Denicol:2012cn,Florkowski:2013lya}.
The time derivative is defined as $\dot{A}=DA$ with $D=u^\mu \partial_\mu$.
Because of the relaxation time terms, IS hydrodynamics describes the non-equilibrium dynamics beyond the Navier-Stokes hydrodynamics.

\subsubsection{Linear response functions}
Using Eqs.~\eqref{eq:NS} and \eqref{eq:IS}, we would like to linearize the equations in order to obtain the linear response functions.
The formulation of the linear response in hydrodynamics is well explained in Ref.~\cite{Kovtun:2012rj}. 
If we consider a fluid near static equilibrium,
the energy density and the fluid velocity can be written as,
\begin{align}
\label{eq:perturbed_energy}
  e(t,x) &= e_0 + \delta e(t,x), \\
\label{eq:perturbed_velocity}
  u^\mu(t,x) &= (1,\vec{0}) + \delta u^\mu(t,x),
\end{align}
where $\delta u^\mu = (0, \delta \vec{v}(t,x))$.
The perturbation is chosen such that its wave vector lies along the $x$ axis, i.e., $\vec{k}=(k,0,0)$.
By inserting Eqs.~\eqref{eq:perturbed_energy} and \eqref{eq:perturbed_velocity} into Eq.~\eqref{eq:hydroEq} and linearizing the equations, 
one can obtain the equation of motion for the perturbation in $k$-space in a general form as: 
\begin{align}
\label{eq:Kovtun2.10}
\partial_t  (\Phi_{a} (t,k)) + M_{ab}(k) (\Phi_{b}(t,k)) = 0, 
\end{align}
where $\Phi$ is an ($n\times 1$) matrix of $n$ hydrodynamic variables such as \{$\delta e, \delta u, \cdots$\}.
By performing the Laplace transform on Eq.~\eqref{eq:Kovtun2.10}, and defining $K_{ab} = - i\omega \delta_{ab} +M_{ab}$, the equation becomes,
\begin{align}
\label{eq:initialvalue_response_general}
        \Phi_{a} (\omega, k)  = (K^{-1})_{ab} \Phi^0_{b} (k),
\end{align}
with $\Phi^0_{a}(k)= \Phi_a(t=0,k)$. 
Therefore, $(K^{-1})_{ab}$ gives the matrix of linear responses to the perturbations to the system at an initial time, which we call
initial-value response functions.

In the case where the constitutive equation of viscous terms is given by the Navier-Stokes formalism \eqref{eq:NS},  $\Phi_a$ and $K$ in Eq.~\eqref{eq:initialvalue_response_general} in the sound channel become: 
\begin{align}
\label{eq:matrixNS}
    \Phi =
    \begin{pmatrix}
        \delta e \\
        \delta u_x
    \end{pmatrix}
    ,\quad
    K=
    \begin{pmatrix}
    -i\omega & ikw \\
    ikc_s^2/w & -i\omega + \Gamma_sk^2,
    \end{pmatrix}
\end{align}
with attenuation length $\Gamma_s = 4\eta/(3w)$ and enthalpy $w=e + P =sT$. From Eq.~\eqref{eq:initialvalue_response_general}, 
if we choose an initial energy-density perturbation with \(\delta u_x =0\), 
the energy-energy initial-value response function is obtained from $(K^{-1})_{00}=\delta \tilde{e}_k(\omega)/\delta \tilde{e}_k^0$,
\begin{align}
\label{eq:IVresponse_NS}
\cfrac{\delta \tilde{e}_k(\omega)}{\delta \tilde{e}_k^0} 
= \cfrac{i\omega - \Gamma_s k^2}{\omega^2 + i\omega \Gamma_s k^2 - c_s^2k^2},
\end{align}
where $\tilde{e}_k(\omega) = \Phi_{\delta e}(\omega)$.


Similarly, in the case where the constitutive equation of viscous terms is given by the Israel-Stewart (IS) equations Eq.~\eqref{eq:IS}, 
$\Phi_a$ and $K$ in Eq.~\eqref{eq:initialvalue_response_general} in the sound channel become: 
\begin{align}
\label{eq:matrixIS}
    \Phi =
    \begin{pmatrix}
        \delta e \\
        \delta u_x \\
        \ \pi^{xx}
    \end{pmatrix}
    ,\quad
    K=
    \begin{pmatrix}
    -i\omega & ikw & 0 \\
    ikc_s^2/w & -i\omega & ik/w \\
    0 & iw\Gamma_s k/\tau_{\pi}   & \tau_\pi^{-1}-i\omega
    \end{pmatrix},
\end{align}
and,
\begin{align}
\label{eq:IVresponse_IS}
    \cfrac{\delta \tilde{e}_k(\omega)}{\delta \tilde{e}_k^0} 
= \cfrac{-i \Gamma_s k^2 + \omega (-1 +i\omega \tau_\pi)}{ic_s^2 k^2 (-1 + i\omega \tau_\pi ) + i\omega (i\Gamma_s k^2 + \omega - i\omega^2 \tau_\pi)}.
\end{align}
Both initial response functions in Eqs.~\eqref{eq:IVresponse_IS} and \eqref{eq:IVresponse_NS}
contain the same hydrodynamic poles at small $k$.
One of the main differences between the linear response functions in Eqs.~\eqref{eq:IVresponse_IS} and \eqref{eq:IVresponse_NS} is that Eq.~\eqref{eq:IVresponse_IS} contains a non-hydrodynamic pole.
We use these analytical results of the initial response functions as a reference for simulation results in Sec.~\ref{subsec:extracted_poles}.

\subsection{Transport}
\subsubsection{Relaxation time approximation}

The Boltzmann equation is a microscopic theory that describes the space-time evolution of the phase-space distribution of each particle species $f^i(t, x,p)$.  
\begin{align}
    p^\mu\partial_\mu f^i(t,x,p) = \mathcal{C}_{1\leftrightarrow 2}[f^i] + \mathcal{C}_{2\leftrightarrow 2}[f^i] \cdots.
\end{align}
%
The relaxation-time approximation (RTA) replaces the above full collision integral with:
\begin{align}
\label{eq:RTAcollisionterm}
\mathcal{C}[f]
\Rightarrow
-\frac{p\cdot u}{\tau_R}
\left(f-f_{\rm eq}\right),
\end{align}
where $\tau_R$ is the relaxation time of the system.
Its main advantage is that it preserves the essential description of the Boltzmann equation while making the equation analytically tractable.

\subsubsection{Linear response from RTA}
The initial-value linear response function of Boltzmann equations with RTA was obtained in Ref.~\cite{Ochsenfeld:2023wxz}.
Equation~\eqref{eq:RTAcollisionterm} is linearized by considering a small perturbation of $f(t, x,p)$ on a global homogeneous equilibrium background $f_0(p)$, $i.e. \ f(t,x,p) = f_0(p) +\delta f(t,x,p)$. Similarly, $f_{\mathrm{eq}} (t,x,p) = f_0 (p)+\delta f_{\mathrm{eq}} (t,x,p)$ where $f_{\mathrm{eq}} = \exp(-p\cdot u /T)$.
Choosing a perturbation with a wave vector $\vec{k} = (0,0,k)$ and defining $\mu \equiv p^z/E = \cos\theta $, the linearized Boltzmann equation with RTA in $k$-space becomes,
\begin{align}
\label{eq:linearizedRTA}
    (\partial_t + i\mu k) \delta \tilde{f}_k = - \tau_R^{-1} (\delta \tilde{f}_k - \delta \tilde{f}_{k,\mathrm{eq}}),
\end{align}
where $\delta \tilde{f}_k = \delta \tilde{f}_k (t ,k,p)$ is a spatial Fourier transform of $\delta f(t,x,p)$. 

The Fourier transform of the perturbed equilibrium distributions $\delta \tilde{f}_{k, \rm eq}$ is obtained by expanding $f_{\mathrm{eq}} = \exp(-p\cdot u /T)$ with,
\begin{align}
    T&= T_0 + \delta T \\
    u^\mu& = (1,\vec{0}) + \delta u^\mu,
\end{align} 
and by performing a spatial Fourier transform on the obtained $\delta f_{\rm eq}$.
The $\delta \tilde{T}_k$ and $\delta \tilde{u}^z_k$ in $\delta \tilde{f}_{k,\rm eq}$ 
can be expressed in terms of $\delta \tilde{e}_k$
and $\delta \tilde{T}^{tz}$
by using conformality, and by choosing the Landau frame.

By using the Laplace transform of Eq.~\eqref{eq:linearizedRTA}, the obtained $\delta \tilde{f}_{k,\rm eq}$, and the moments of $\delta \tilde{f}_k$:
\begin{align}
    \delta \tilde{e}_k = g \int \cfrac{d^3p}{(2\pi)^{3}} E\delta \tilde{f}_k, \quad 
    \delta \tilde{T}^{tz}_k = g \int \cfrac{d^3p}{(2\pi)^3} \mu E \delta \tilde{f}_k,
\end{align}
the initial-value response function of the Boltzmann equations with RTA is obtained,
\begin{align}
    \cfrac{\delta \tilde{e}_k (\omega)}{\delta \tilde{e}_{k}^0}
    = -\cfrac{6ik\tau_R + (3+k^2\tau_R^2 - 3i \omega \tau_R)L}
    {k^2\tau_RL + 2ik^3\tau_R^2 - 6k\tau_R \omega + 3\omega(i+\omega\tau_R)L},
    \label{eq:IVresponse_RTA}
\end{align}
where 
\begin{align}
    L = \ln \left(\cfrac{1-i\tau_R(k+\omega)}{1+i\tau_R(k-\omega)}\right).
\end{align}
Therefore, in addition to the hydrodynamic poles, Eq.~\eqref{eq:IVresponse_RTA} has a logarithmic branch-cut whose contribution becomes large with a small wavelength \cite{Romatschke:2015gic}.
The branch-cut appears as a result of the free-streaming massless particles in a continuous range of directions.
As with Eqs.~\eqref{eq:IVresponse_NS} and \eqref{eq:IVresponse_IS}, we use this analytical result of the initial response function as a reference for simulation results in Sec.~\ref{subsec:extracted_poles}.

\subsection{Initial-value response function and dispersion relation}
To extract the pole locations from the simulations of real-time dynamics, it is more convenient to know the initial-value response function as a function of time.
Before we proceed, we briefly review the relation between the response functions that we obtained in the previous section and their poles. 
In general, if only the initial energy perturbation, $\delta \tilde{e}_k(0)$, is excited and  the initial velocity is set to zero $\delta u^\mu (t_0, x) = 0$, the time evolution of a small perturbation $\delta e_k$ can be written as,
\begin{align}
\label{eq:deltae_kt}
    \delta \tilde{e}_k(t) = 
\tilde{G}_{ee, k}^{\mathrm{ival}}(t) \delta \tilde{e}_k (0),
\end{align}
where $\tilde{G}^{\rm ival}_{ee,k}(t)$ is the initial-value response functions,
and,
\begin{align}
\label{eq:Gival_FT}
\delta \tilde{e}_k(t) /\delta \tilde{e}_k(0)
    & = \tilde{G}_{ee,k}^{\mathrm{ival}}(t) \nonumber \\
    & = \frac{1}{2\pi} \int_{-\infty}^{+\infty} d\omega  \ e^{-i\omega t} \tilde{G}_{ee,k}^{\mathrm{ival}}(\omega) \nonumber \\
    & = \sum_{i=1}^{n_p} A_i(k) e^{-i\omega_i^{p} t}.
\end{align}
The relation from the second line to the third holds if we assume that 
$\tilde{G}_{ee,k}^{\mathrm{ival}}(\omega)$ has $n_p$ simple poles $\omega = \omega^p_i$. The $A_i(k)$ is a coefficient determined by the residue at the corresponding pole.

In the long-wavelength limit, the sound channel has two poles: $\omega^{p}_{\pm} = \pm \Omega^{p} - i \Gamma^{p}$ with $\Omega^{p}(k) = |\Ree{\omega^{p}(k)}|$ and $\Gamma^{p}(k) = |\Imm{\omega^{p}}(k)|$.
For the perturbation that is real,
Eq.~\eqref{eq:Gival_FT} becomes
\begin{align}
\label{eq:FittingFunction}
\delta \tilde{e}_k(t) /\delta \tilde{e}_k(0)
   = A e^{-\Gamma^{p}t}[\cos(\Omega^{p}t + \phi )],
\end{align}
where $\phi$ is a phase, and $A$ is a proportionality coefficient.
The imaginary part of the pole characterizes damping, while the real part characterizes frequency.
Therefore, the pole locations can be extracted from the hydrodynamic and transport simulations by fitting the evolution of the energy density perturbation with Eq.~\eqref{eq:FittingFunction}.

To find the locations of the poles in the initial-value response functions, 
the denominators of Eqs.~\eqref{eq:IVresponse_NS}, \eqref{eq:IVresponse_IS}, and \eqref{eq:IVresponse_RTA} are set to zero.
\footnote{
Note that, in principle, the initial-value response function and the retarded Green's function are different. However, as explained in Refs.~\cite{Kovtun:2012rj,Ke:2023bei}, the analytical structure including the locations of poles is the same in both.
}
In the small $k$ limit, the hydrodynamic poles from NS, IS, and RTA converge to,
\begin{align}
\label{eq:dispersion_relation_NS}
\omega(k)=\pm c_sk - i(\Gamma_s/2)k^2 + \mathcal{O}(k^3).
\end{align}

\section{Methods}
\label{sec:methods}
One of the main goals of this work is to establish a method for extracting poles of linear response functions from numerical simulations commonly used in model-to-data comparisons for relativistic heavy-ion collisions.

We describe the numerical setups for viscous hydrodynamic and Boltzmann transport simulations in Sec.~\ref{subsec:Numerical_implementations} and the method for extracting the linear response from these simulations in Sec.~\ref{subsec:Extraction_of_initial_value_response_function}.

\subsection{Numerical implementations}
\label{subsec:Numerical_implementations}
\subsubsection{Hydrodynamic simulations with finite difference method}
In this work, we use relativistic hydrodynamics in
(1+1)D Cartesian coordinates, which should be sufficient for the scope of this paper.
The accuracy of the simulation is important in order to study response functions because the numerical viscosity can introduce artificial excessive diffusion depending on the numerical scheme.
One of the most common methods used for relativistic heavy-ion collisions to solve the equations is the finite difference method (FDM). 
\footnote{For a smoothed-particle hydrodynamics (SPH), a Lagrangian particle method, see, for instance, \cite{Pala:2025qoa}.}
In FDM, the accuracy depends on the discretization methods in space and time, and reconstruction of the conserved quantity from piecewise discretized hydrodynamic fields.

In this study, we mainly use PPM+HLLE \cite{Tachibana:2014yai,Tachibana:2014lja,Murase:2015oie} for the space discretization method.
The piecewise parabolic method (PPM) is a higher-order extension of Godunov's method where a continuous conserved current is reconstructed by a parabolic function within a single fluid cell \cite{Colella:1982ee}.
The Harten-Lax-van-Leer-Einfeldt (HLLE) solver uses the conserved current at cell interfaces obtained by PPM and calculates fluid fluxes \cite{Rischke:1998fq,Amiram:2006zjz,Einfeldt,Karpenko:2013wva}.
Note that the PPM+HLLE scheme is adapted from Ref.~\cite{PPMHLLE}. We modified the code to implement the second order viscous hydrodynamics in (1+1)D Cartesian coordinates based on Ref.~\cite{Molnar:2009tx}.
For time discretization, we used an explicit two-stage predictor–corrector scheme, which is expected to have second-order accuracy in time \cite{Molnar:2009tx,Karpenko:2013wva}.
Also, the periodic boundary condition is imposed.

\subsubsection{Boltzmann Transport with geometrical interpretation of cross section}
\label{subsub:transport_with_geometrical_interpretation_of_cross_section}
The RTA makes it possible to analytically solve the equations. On the other hand, 
direct Monte Carlo simulations of the Boltzmann equation are necessary to describe experimental data that result from the complex dynamics of quarks and gluons or hadrons in relativistic heavy-ion collisions.
In the direct Monte-Carlo simulations, the microscopic interaction is performed by cascades of collisions of ``classical'' particles with a position and momentum expressed by delta functions. The interaction of quarks, gluons or hadrons is encoded through their total and differential cross sections based on perturbative QCD or measured properties of hadrons from which the collision probability and outgoing momenta are sampled.

We perform the direct Monte-Carlo simulation of the Boltzmann equation in 3D Cartesian coordinates in a periodic box, using JAM2 \cite{Nara:2023vrq,JAM2,Nara:2019crj} in which Poincaré invariance is introduced in the equations of motion of particles.
In this work, we consider only massless gluons as the only degrees of freedom with an isotropic cross section $\sigma_{gg \leftrightarrow gg} = 2.6$ mb.
Momenta of initial gluons are sampled from a massless Boltzmann distribution with degeneracy $g=16$, while coordinates are sampled so that a given initial energy profile is realized. We will discuss the initial energy profile in the next section.

In the direct simulation of the Boltzmann equations, collisions of particles take place at a distance $r_p \le \sqrt{\sigma/\pi}$. Therefore, the description stays reasonable only when $l_p \gg r_p$ where $l_p = \rho^{-1/3}$ is an estimate of the distance between particles, and $\rho$ is a the number density of the system.
A conventional method used in transport simulations to overcome this limitation is the test particle method \cite{Zhang:1998tj,Molnar:2000jh}.
Under this method, the cross section $\sigma$ and the number density are rescaled as:
\begin{align}
    \sigma \rightarrow \sigma /N_{\rm test}, \quad
    \rho \rightarrow \rho N_{\rm test},
\end{align}
so that the corresponding mean free path $\lambda_{\rm mfp} = 1/(\rho \sigma)$ stays the same.
The $N_{\mathrm{test}}$ is chosen to satisfy $N_{\mathrm{test}} \gg  (\sigma/\pi)^3 \rho^2$ so that the relation $r_p/l_p\ll 1$ holds since $r_p/l_p$ shrinks by $N_{\rm test}^{-1/6}$. 
In principle, increasing $N_{\rm test}$
reduces numerical sampling noise, yielding smoother estimates of the observables even from a single transport run of the transport.
However, we need to choose $N_{\rm test}$ so that a single run of the transport still fits within the wall time of the computational resource, which limits the maximum of $N_{\rm test}$ with the current implementation of the code. Therefore, we still run the transport $10^2$-$10^3$ times in this study even with the test particle method to obtain a smooth profile of the observables.

\subsection{Extraction of initial-value response function}
\label{subsec:Extraction_of_initial_value_response_function}

In order to study the linear response functions from fully non-linear hydrodynamic and transport simulations, we prepare an initial energy density profile with a small perturbation on a uniform background $e_0$ in a periodic box as,
\begin{align}
\label{eq:energy_density_initial_condition}
    e(t_0,x) = e_0 + \delta e(t_0, x),
\end{align}
where $t_0 =0$ fm is the initial time, and 
\begin{align}
\label{eq:delta_e_t0}
    \delta e(t_0,x) = A_0 e_0 \cos(kx),
\end{align}
where $A_0$ is an amplitude which is set to $0.1$ throughout this work, $k =2\pi/\lambda$ is a wavenumber, and $\lambda$ is the wavelength of the initial perturbation. 
The perturbation is added only in the $x$-direction in the simulation space $[-L/2,L/2]$.
The simulation range $L$ is adjusted to be an integer multiple of the wavelength. 
Note that the initial velocity is set to zero: $\delta u^\mu (t_0, x) = 0$.

\begin{figure}[h]
    \centering
    \includegraphics[width=1.0\linewidth]{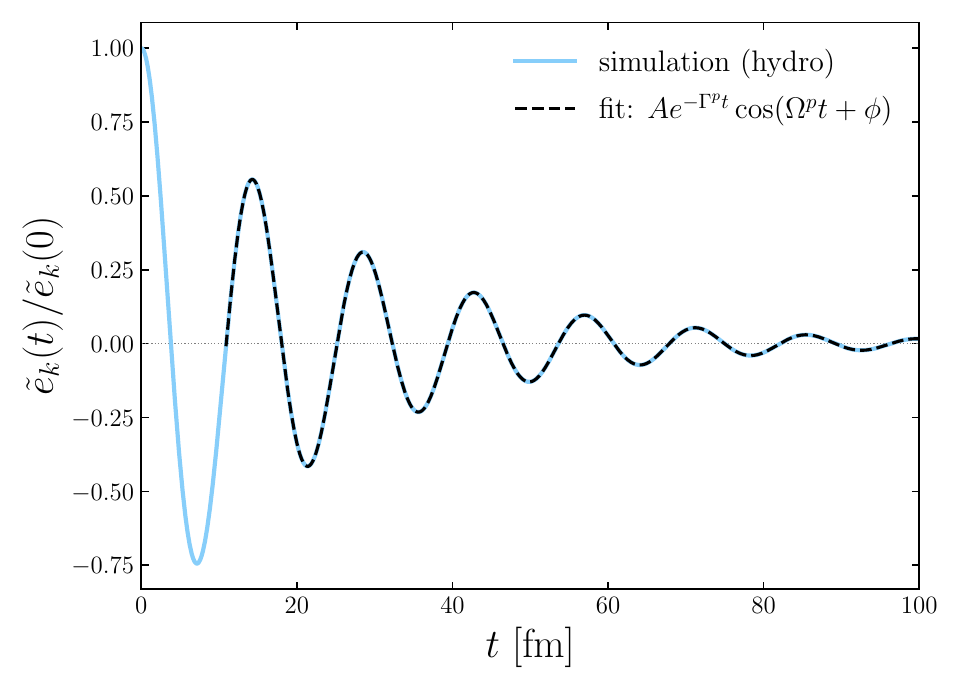}
    \caption{$\delta \tilde{e}_k (t)$ obtained from a hydrodynamic simulation (solid) and the fitted result (dashed). The parameters used for this result are $\eta/s=0.2$, $T_0= 0.35$ GeV, and $k=0.75$ $\mathrm{fm}^{-1}$.
    The obtained fitting parameters are:
    $A = 1.00, \Gamma^p = 0.041~\text{fm}^{-1}, \Omega^p = 0.44~\text{fm}^{-1}, \phi       = - 0.10~\text{rad}$. 
    }
    \label{fig:nqt_hydro}
\end{figure}

We run simulations up to $t_{\mathrm{max}}$, 
and obtain $\delta e (t,x)$ for $0<t<t_{\mathrm{max}}$.
In hydrodynamic simulations, $t_{\mathrm{max}} = 260$ fm is chosen so that it covers the time range until the envelope of the perturbation damps to at least less than 10\% of its initial amplitude for most cases.
For some values of $\bar{k}$, the damping is too weak for the perturbation to decay sufficiently within a practical simulation time. We therefore stop these simulations at a finite time $t_{\mathrm{max}}$ and fit the amplitude before the perturbation has fully decayed. We verified that $t_{\mathrm{max}} = 260\,\mathrm{fm}$ provides a sufficiently long fitting window, as further increasing $t_{\mathrm{max}}$ does not significantly change the fit results.

On the other hand, $t_{\mathrm{max}}=100$ fm is chosen in transport simulations. The initial perturbation almost dies out and statistical noise becomes dominant by $t_{\mathrm{max}}=100$ fm for all the parameter sets presented below.
As discussed in Sec.~\ref{subsub:transport_with_geometrical_interpretation_of_cross_section}, we simulate Boltzmann transport in 3D using a particle description. To prepare initial conditions consistent with those of hydrodynamics,
we sample the initial gluon momenta from a massless Boltzmann distribution, with directions sampled uniformly over solid angle so that no direction is preferred. We sample their positions to reproduce the initial energy density profile defined in Eq.~\eqref{eq:energy_density_initial_condition}.
The resulting energy density varies along the $x$ direction and is uniform in the $y$ and $z$ directions.
We calculate the one-dimensional energy-density profile \(e(t,x)_{\rm run}\) from each transport run by binning the particles along the \(x\) direction and averaging it over the full \(y\)- and \(z\)-ranges of the periodic box. Then, \(e(t,x)_{\rm run}\) is averaged over runs to obtain a smooth profile.
To obtain \(\delta e(t,x)\), the global equilibrium energy density $e_0$ is reconstructed by averaging $e(t,x)$ over 
$[-L/2,L/2]$ at the last time step when it is expected to be close to the global equilibrium. We checked that the obtained $e_0$ is consistent with the energy density expected from a massless Boltzmann distribution with initial temperature $T_0$.

After we obtain $\delta e (t,x)$ from simulations,
the discrete spatial Fourier transform is calculated over $[-\lambda/2, \lambda/2]$ with space step $\Delta x$ for the time windows $t_{\rm min} <t<t_{\rm max}$ for the hydrodynamic simulations and $t_{\rm min} <t<t_{\rm cut}$ for the transport simulations using:
\begin{equation}
  \delta \tilde{e}_k(t) = \Delta x \sum_{j} \delta e(t, x_j)\,e^{-i k x_j},
  \label{eq:DFT}
\end{equation}
where $x_j = x_{j-1} + \Delta x$. 
The $\Delta x$ is the cell width in the results from hydrodynamic simulation, and the binning width of the energy density distribution for transport,
which are $\Delta x = 0.02 \ \rm fm$ 
\footnote{
When necessary, the cell width is adjusted by less than 10\% of its original value so that the box length equals an integer number of wavelengths.}
and $\Delta x = 0.15 \ \rm fm$, respectively. 
We introduce the starting time of the fitting window $t_{\rm min}$ based on the discussion in Ref.~\cite{Du:2023bwi}.
Throughout this study, $t_{\rm min}$ is fixed to $c_s k t_{\rm min} = 3\pi/2$. In this way, we avoid 
the effect of non-hydrodynamic mode \cite{Du:2023bwi} and the numerical inaccuracies which are present at early times.

Note that, only for the results obtained from transport simulations, we also impose an upper time limit $t_{\rm cut}<t_{\max}$ on the fitting window to exclude late-time statistical noise that may bias the fit.
For hydrodynamic simulations, the fitting window extends from $t_{\rm min}$ to $t_{\rm max}$.
See Appendix \ref{appendix:fittings_of_deltae} for the $t_{\rm min}$ dependence of the fitting results, and for the methodology of $t_{\rm cut}$ estimation.

\begin{figure}[h]
    \centering
    \includegraphics[width=1.0\linewidth]{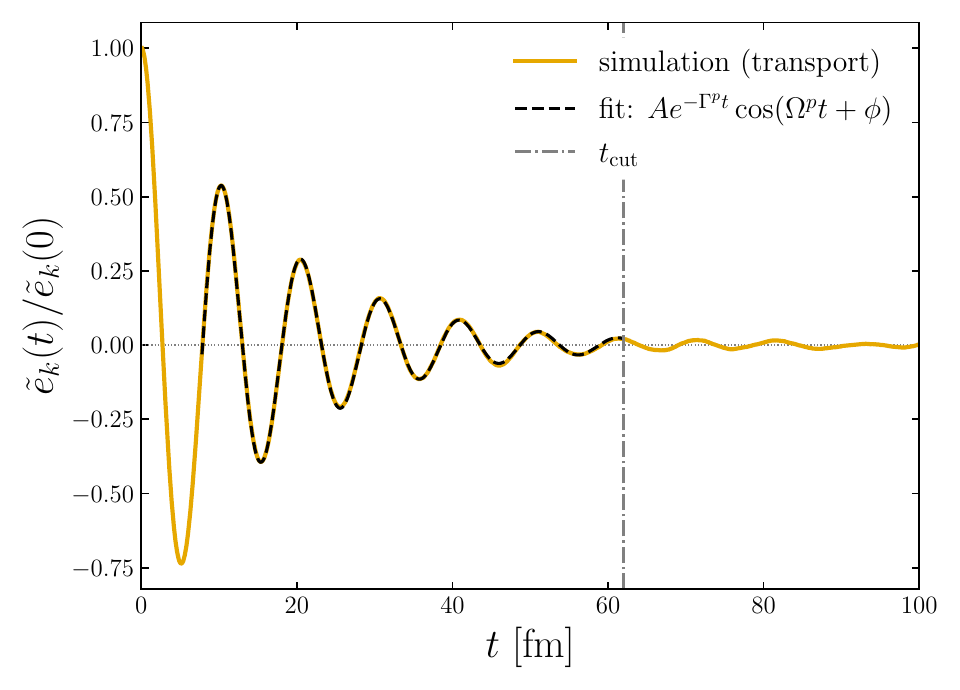}
    \caption{$\delta \tilde{e}_k (t)$ obtained from a transport simulation (solid) and fitted result (dashed). The parameter set used in this result are, $T_0= 0.45$ GeV, and $k=1.047$ $\mathrm{fm}^{-1}$.
    Extracted values are:
    $A     = 1.01, \Gamma^p = 0.061~\text{fm}^{-1},
    \Omega^p = 0.62~\text{fm}^{-1}, 
    \phi = -0.14~\text{rad}$. The maximum time of the fitting window is $t_{\rm cut}\sim 62$ fm.}
    \label{fig:nqt_transport}
\end{figure}

We extract $\Omega^{p}$ and $\Gamma^{p}$ by fitting $\delta \tilde{e}_k(t) /\delta \tilde{e}_k(0)$ with Eq.~\eqref{eq:FittingFunction}.
Figures~\ref{fig:nqt_hydro} and \ref{fig:nqt_transport} show examples of $\delta \tilde{e}_k(t) /\delta \tilde{e}_k(0)$ obtained from hydrodynamics and transport simulations for the given parameters 
(See the captions for the parameter sets used).
For these particular choices of parameters,
in both cases of Figs.~\ref{fig:nqt_hydro} and \ref{fig:nqt_transport},
the obtained initial-value response functions are well fitted by Eq.~\eqref{eq:FittingFunction} that assumes only the hydrodynamic sound mode. This means that the linear response functions in the fitting window are dominated by the hydrodynamic sound modes.

In general, the wavenumber $k$ and the attenuation length $\Gamma_s$ are the two principal variables that determine $\Ree \omega$ and $\Imm \omega$ in our case. In hydrodynamics, the attenuation length is controlled by $\eta/s$ and temperature.
In our transport with a constant isotropic cross section, the attenuation length is controlled by temperature and cross section $\sigma$. 
In this study, we vary $\eta/s$ in hydrodynamics, and the equilibrium temperature $T_0$ in transport keeping the others constant.
\footnote{
We vary the temperature rather than the cross section in transport simulations because our primary quantity of interest, \(\eta/s \propto T^{-2}\sigma^{-1}\), depends more strongly on temperature. This allows us to explore a wider range of \(\eta/s\).
}
Although one could vary other sets of parameters, the scaling discussed in the next section makes such variations equivalent to varying the parameters listed above.

\section{Results}
\label{sec:results}
We perform hydrodynamic simulations for five wavenumbers ranging from 0.5 to 3.142 $\rm fm^{-1}$, and eight values of $\eta/s$ ranging from 0.025 to 0.2. Simulating every combination of these parameters yields a total of 40 data points presented below.
These values of the wavenumbers are chosen either by fixing the box length $L$ and selecting $k=2\pi n/L$, or by fixing $k$ and setting $L=2\pi n/k$, where $n$ is a positive integer. In both cases, the box contains an integer number of wavelengths, as required by the periodic boundary conditions.
Throughout this study, we use $e_0 = 10 \ \rm GeV/fm^3$, which corresponds to $T_0\sim 0.35 \ \rm GeV$.

We perform transport simulations for three wavenumbers, \(k = 1.047, 2.094,\) and \(3.142\,\mathrm{fm}^{-1}\), and three temperatures \(T=0.25,0.35\) and \(0.45\, \rm GeV\). 
The volume of a periodic box is set to $(6\times6\times6) \ \rm fm^{3}= 216 \ \rm fm^{3}$.
The shear viscosity per entropy density $\eta/s$ used to compute $\bar{k}$ (the horizontal axis) for the transport results is calculated as
$\eta/s=\eta\pi^2/(64T^3)$ where $\eta = 1.2T/\sigma_{\rm tot}$ based on Ref.~\cite{Plumari:2012ep}.
The entropy density $s$ is obtained from a massless Boltzmann distribution.
This relation between $\eta$ and $\sigma_{\rm tot}$ is analytically obtained from the RTA of Boltzmann equations, taking into account the difference between the total cross section $\sigma_{\rm tot}$, which is used as the collision criterion in transport, and the transport cross section $\sigma_{\rm tr}$, which is obtained by weighting the differential cross section by $(1-\cos^2\theta)$ to properly account for momentum transfer. The relation was also confirmed with the Chapman-Enskog approach.
By matching second-order hydrodynamics and the RTA, $\tau_R = \tau_\pi$ \cite{Romatschke:2015gic}.

In Ref.~\cite{Du:2023bwi}, 
it was demonstrated that the initial-value response function shows universality over different values of viscosity and temperature when plotted in terms of dimensionless rescaled values.
Based on this work, the results are shown in terms of the following variables:
\begin{align}
    \bar{k} = k\eta/(sT), \quad
    \bar{t} = t (sT)/\eta
\end{align}
and correspondingly,
\begin{align}
\bar{\omega} = \omega \eta/(sT).
\end{align}

\begin{figure}
    \centering
    \includegraphics[width=1.0\linewidth]{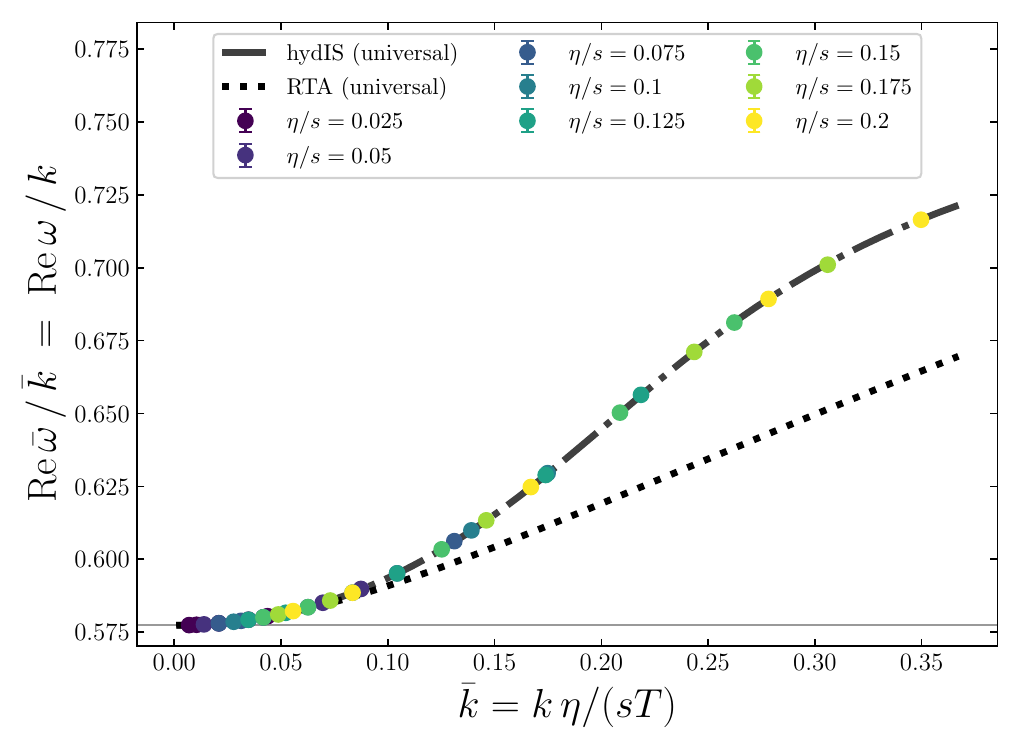}
    \includegraphics[width=1.0\linewidth]{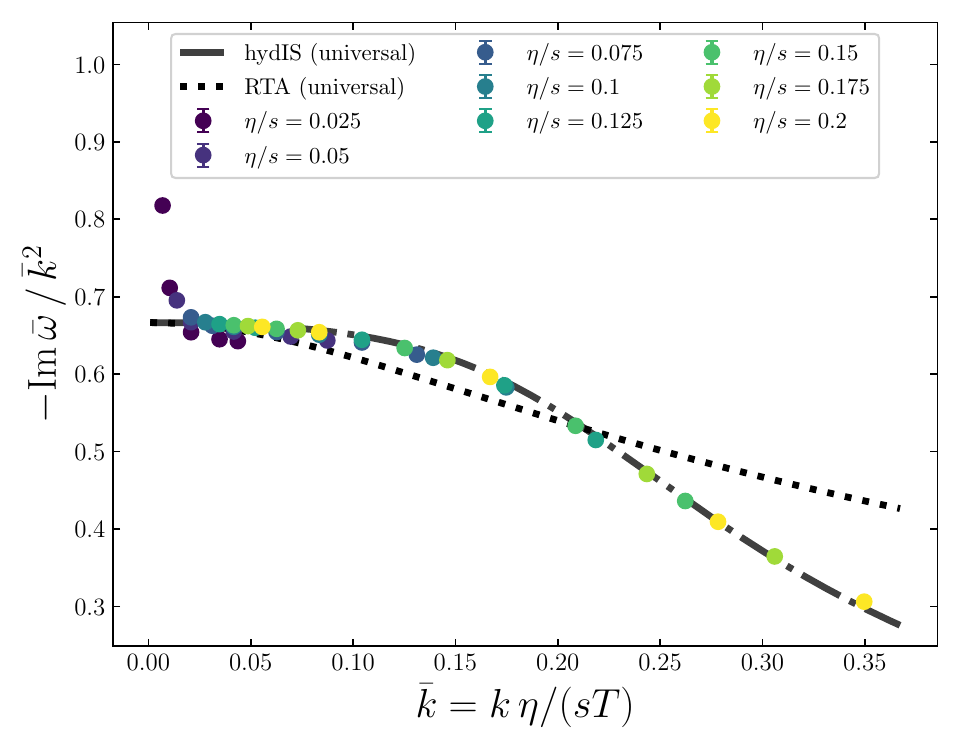}
    \caption{$\Ree \bar{\omega}/\bar{k}$ (top) and $-\Imm \bar{\omega}/\bar{k}^2$ (bottom) as a function of $\bar{k}$ extracted from the IS hydrodynamic simulations (filled circles), compared to the analytical results from the IS (dash-dotted line) and RTA (dotted line). Simulation results with different $\eta/s$ are plotted with different colors in filled circles with fitting uncertainties at the 68\% CL. The gray horizontal line shown in the top figure is an NS-limit reference at small $\bar{k}$, $\Ree \bar{\omega}/\bar{k}= c_s = 1/\sqrt{3}$.
    }
    \label{fig:poles_hydro}
\end{figure}

\begin{figure}
    \centering
    \includegraphics[width=1.0\linewidth]{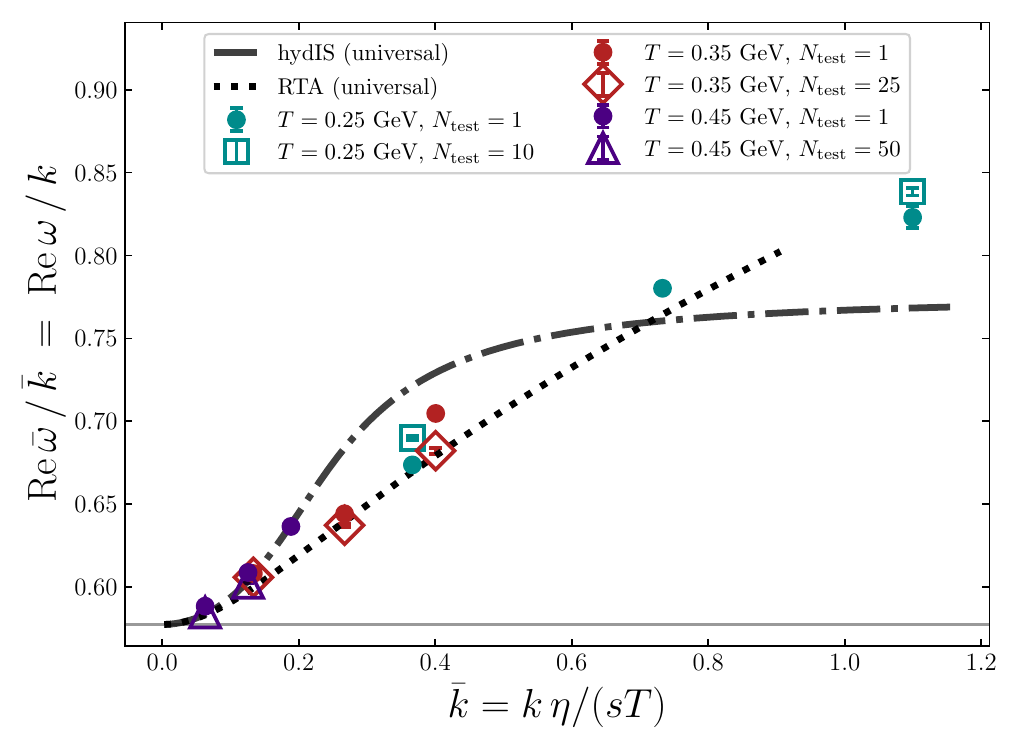}
    \includegraphics[width=1.0\linewidth]{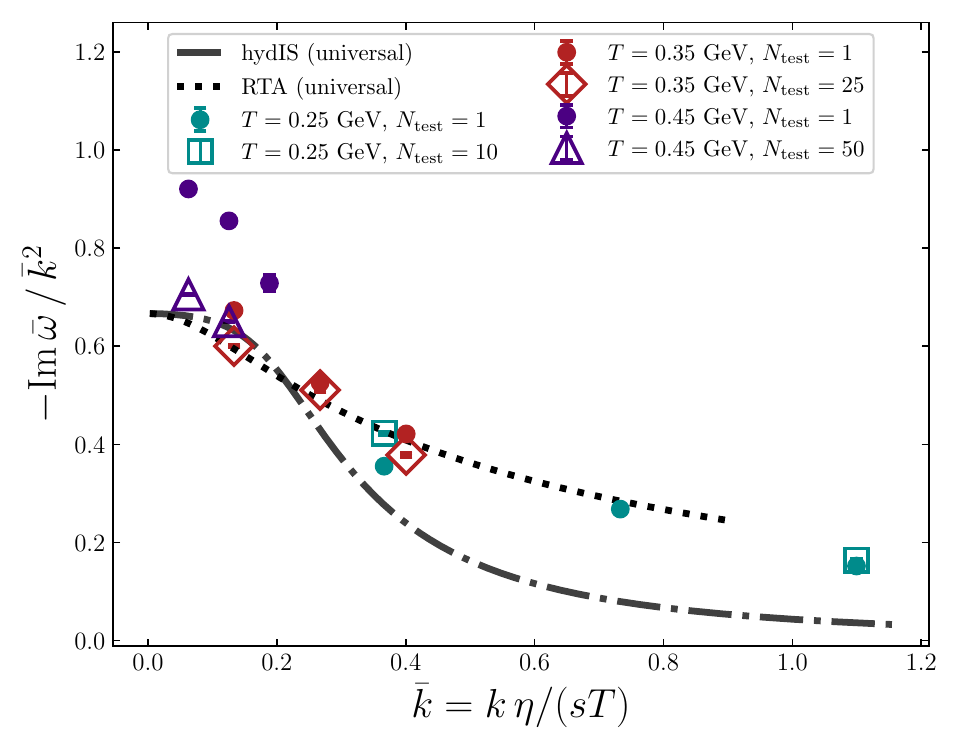}
    \caption{$\Ree \bar{\omega}/\bar{k}$ (top) and $-\Imm \bar{\omega}/\bar{k}^2$ (bottom) as a function of $\bar{k}$ extracted from transport simulations (filled circles and open symbols), compared to the analytical results from IS (dash-dotted line) and RTA (dotted line).
    Simulation results with different $T$ are plotted with different colors. Filled circles (open symbols) are results without (with) the test particle method. 
    Error bars show fitting uncertainties at the 68\% CL. The gray horizontal line in the top panel is the NS-limit reference at small $\bar{k}$, $\Ree \bar{\omega}/\bar{k}= c_s = 1/\sqrt{3}$.
    }
    \label{fig:poles_transport}
\end{figure}

\subsection{Extracted poles}
\label{subsec:extracted_poles}
Figure~\ref{fig:poles_hydro} shows the $\bar{k}$-dependence of $\Ree \bar{\omega}/\bar{k}$ and $-\Imm \bar{\omega}/\bar{k}^2$ obtained by fitting the hydrodynamic simulation results with Eq.~\eqref{eq:FittingFunction} \cite{supplemental}.
The analytical results are obtained using a root-finding method, with the denominators of Eqs.~\eqref{eq:IVresponse_IS} and \eqref{eq:IVresponse_RTA} set equal to zero.
The horizontal axis, $\bar{k}$, is calculated from the input $\eta/s$ and the background temperature $T_0$. 
The extracted poles of the response functions using different values of $\eta/s$ and different wavenumbers $k$ of the initial perturbation show clear scaling with $\bar{k}$. 
Thus, the dimensionless quantity $\bar{k}$ is a good measure of the applicable range of hydrodynamics in terms of the linear response functions.
The results for a given $\eta/s$ spread over a potentially wide $\bar{k}$ range. This shows that the applicability of hydrodynamics in terms of linear response depends not only on $\eta/s$ but also on the wavenumber of the perturbation and temperature. 

The results lie on top of the IS analytical curves except for $-\Imm \bar{\omega}/\bar{k}^2$ at $\bar{k}\sim0$.
The matching between the extracted results and the analytical curves is all expected because we solve IS hydrodynamics with numerical simulations.
Thus, in other words, the comparisons between the extracted and analytical poles can be treated as a numerical test of computational fluid dynamics.
Note that we also performed the same analysis using SHASTA (Sharp and Smooth Transport Algorithm) \cite{Boris:1973tjt,Molnar:2009tx}, which is shown in Appendix~\ref{appendix:SHASTA}. The results using the SHASTA show no major difference from the results with PPM+HLLE, demonstrating the equivalent accuracy of the simulations. 

We believe that the deviation between the numerical results and analytical results at small $\bar{k}$ for $-\Imm \bar{\omega}/\bar{k}^2$ originate from the emergence of non-linear response for the following reason.
We checked the corresponding results by changing the amplitude of the perturbation in Appendix \ref{appendix:amplitude_dependence_in_negative_omega}, and observed that the deviation became smaller with a smaller amplitude while the results at large $\bar{k}$ remained unchanged.
Non-linear responses are visible at small $\bar{k}$, where weak damping allows the induced perturbations to persist.
Further investigation of the non-linear response is left for future work.

Figure~\ref{fig:poles_transport} shows the results corresponding to those in Fig.~\ref{fig:poles_hydro} but from transport simulations.
The effect of the test particle method is shown in the comparisons between the filled circles and open symbols at some $\bar{k}$, which is significant at $T=0.45$ GeV for the $-\Imm \bar{\omega}/\bar{k}^2$.
Notice that the test particle method brings the results much closer to the analytical values.
This is consistent with the expectation because the ratio of the particle size to the interparticle distance in collisions becomes $r_p/l_p \sim 0.8$ at $T=0.45$ GeV, while $r_p/l_p \ll 1$ needs to be satisfied as discussed in Sec.~\ref{subsub:transport_with_geometrical_interpretation_of_cross_section}.

The extracted results are almost on top of the analytical results from RTA. This is expected at $k\sim0$ or $\bar{k}\sim0$ because
it is demonstrated in Ref.~\cite{Plumari:2012ep} that the analytically obtained $\eta/s$ from
RTA of Boltzmann equations for massless gluons with isotropic cross-section agrees with numerically calculated $\eta/s$ from the Green-Kubo formula using a partonic transport model.
We also confirmed that the shear viscosity
inferred from the $- \Imm \bar{\omega}/\bar{k}^2$ at small $\bar{k}$ agrees with that obtained from the Green-Kubo formula in our simulations.
On the other hand, the matching of linear responses between the analytical result from RTA and the transport simulation result at comparatively large $\bar{k}$ is non-trivial. This result demonstrates that the RTA well approximates the response of the elastic Boltzmann transport model up to a certain $\bar{k}$, within the parameter range that we studied.

At large $\bar{k}$, 
contribution from the non-hydrodynamic branch cuts in RTA become more dominant compared to the poles of the hydrodynamic sound mode.
The line of the analytical result is, therefore, plotted up to the point where a root finding method breaks down. 
This shows that the linear responses at the largest $\bar{k}$ cannot be represented merely by the hydrodynamic collective dynamics, but instead the influence of independent-particle-like free-streaming dynamics becomes apparent.

The comparison between the poles of linear responses from hydrodynamics and transport simulations 
reveals $\bar{k}$ at which  
the linear response from hydrodynamics starts to fail to describe that from transport simulations.
The split of $\Ree \bar{\omega}/\bar{k}$ and $-\Imm \bar{\omega}/\bar{k}^2$ between hydrodynamics and transport simulations becomes visible around $\bar{k}\sim 0.1$, which is consistent with the results in Ref.~\cite{Du:2023bwi}.
The damping term, \(-\operatorname{Im}\bar{\omega}/\bar{k}^2\), from IS hydrodynamics shows a relative difference of $5\%$ from the RTA at \(\bar{k}\sim0.1\), and the relative difference grows up to almost $80\%$ at \(\bar{k}\sim 0.9\). The deviation in \(\operatorname{Re}\bar{\omega}/\bar{k}\) increases more gradually, reaching only about \(5\%\) at \(\bar{k}\sim 0.9\).
Therefore, the damping part, \(-\operatorname{Im}\bar{\omega}/\bar{k}^2\), is more sensitive to \(\bar{k}\), and we set \(\bar{k}\sim 0.1\) based on the agreement between RTA and IS for this damping part.

\begin{figure}[htpb]
    \centering
    \includegraphics[width=1.0\linewidth]{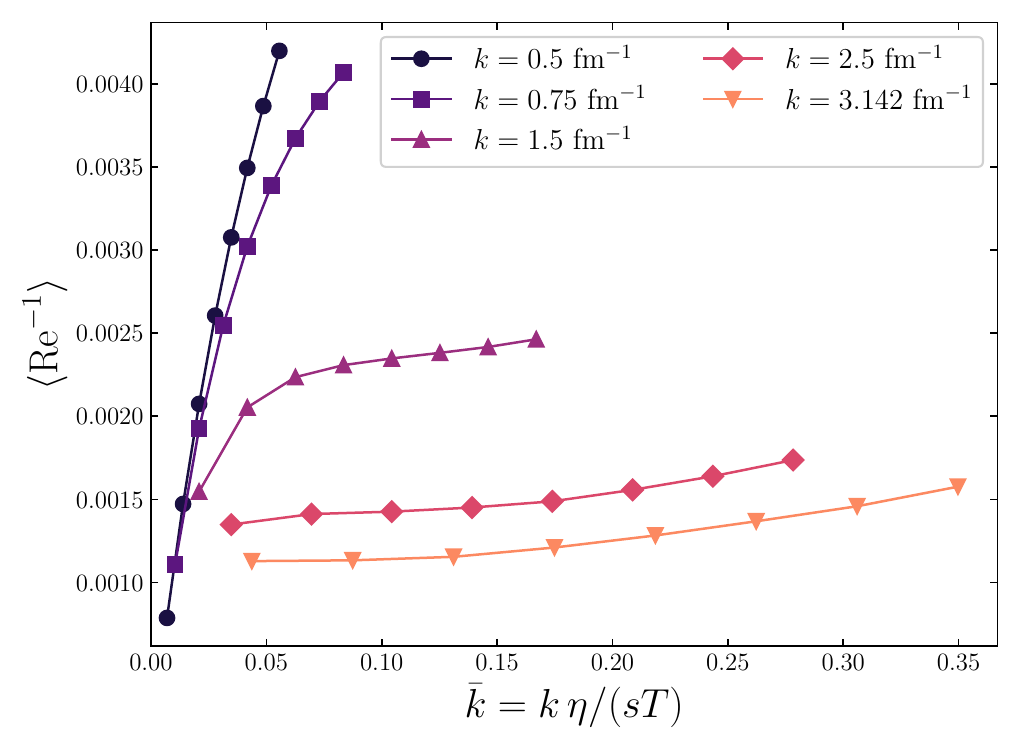}
    \includegraphics[width=1.0\linewidth]{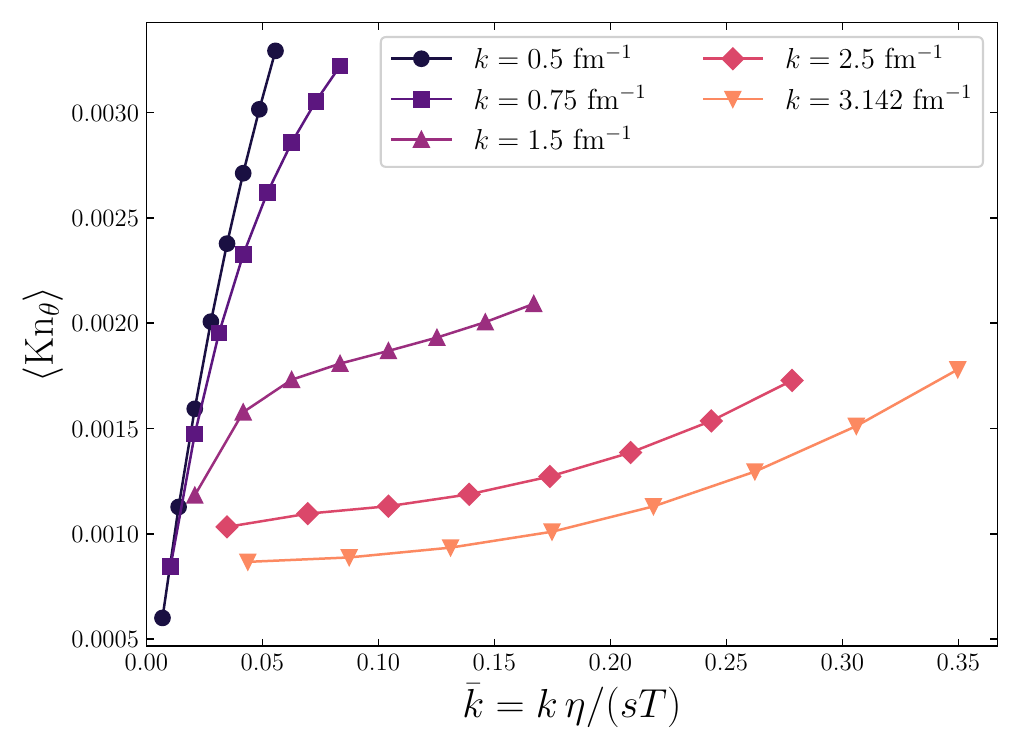}
    \caption{(Top) The averaged inverse Reynolds number, and (bottom) the averaged Knudsen number as functions of $\bar{k}$. Results for different wavenumbers are shown with different filled markers.
    The time average is taken over $0\leq t \leq 50$ fm.
    }
    \label{fig:invRe_Kn_sigma_Kn_theta}
\end{figure}

\subsection{Dependence of the Reynolds and Knudsen numbers on $\bar{k}$}
\label{subsec:DimensionLess_numbers_in_hydrodynamics}
In the previous subsection, the clear scaling of the poles of the linear response on $\bar{k}$ has been demonstrated, which can be interpreted that $\bar{k}$ is a good measure of the applicability of hydrodynamics in terms of linear response.
However, the wavenumber of a perturbation is generally not known or well-defined in realistic simulations.
Therefore, finding dimensionless variables that manifest $\bar{k}$ and can be calculated from hydrodynamic variables without prior knowledge of the wavenumber of a perturbation would be useful for evaluating the applicability of hydrodynamics in simulations for relativistic heavy-ion collisions.

Based on Refs.~\cite{Denicol:2012cn,Niemi:2014wta}, we investigate the $\bar{k}$ dependence of several dimensionless numbers for relativistic hydrodynamics.
Figure~\ref{fig:invRe_Kn_sigma_Kn_theta} shows results of space-time averaged $\langle \Rey^{-1} \rangle$ and $\langle \Kn_\theta \rangle$,
\footnote{
We chose $\Kn_\theta$ for evaluation in this study, as it is mentioned in~\cite{Niemi:2014wta} that 
the Knudsen number based on the expansion rate gives the tightest limit.
}
which are defined as:
\begin{align}
    \Rey^{-1} = \frac{\sqrt{\pi^{\mu\nu} \pi_{\mu\nu} }}{P}, \quad
        \Kn_\theta = \tau_\pi |\theta|,
    \label{eq:invRe_Knudsen}
\end{align}
where $\theta=\partial_\mu u^\mu$ is the expansion rate and $\tau_\pi$ is the relaxation time introduced in Eq.~\eqref{eq:IS}.
The angle brackets $\langle . \rangle$ stand for the space-time average.
The space average is taken over the interval $[-\lambda/2, \lambda/2]$, while the time average is taken over
$0\leq t \leq 50$ fm.
As mentioned above, we would like to investigate whether the Reynolds or Knuden number in Eq.~\eqref{eq:invRe_Knudsen} can estimate $\bar{k}$ without prior knowledge of the wavenumber of a perturbation.
Therefore, we simply fix the time-average window for all parameter sets.
Note that the time-average window starts at $0$ fm instead of $t_{\rm min}= 3\pi/(2c_sk)$ because $t_{\rm min}$ depends on $k$.
We checked that there is no significant quantitative difference between the results obtained from $0\leq t \leq 50$ fm and $t_{\rm min}\leq t \leq 50$ fm.
The dependence of the choice of time-average window is summarized in Appendix~\ref{appendix:time_average_window_dependence_of_Kntheta_and_Knstar}.

First of all, notice in Fig.~\ref{fig:invRe_Kn_sigma_Kn_theta} that the magnitudes of the inverse Reynolds number and the Knudsen number are $\sim \mathcal{O}(10^{-3})$, which is much smaller than the typical values obtained in the simulations of relativistic heavy-ion collisions \cite{Denicol:2012cn,Niemi:2014wta}. Despite the small inverse Reynolds number, the linear response study shows that the applicability of hydrodynamics in terms of linear response functions is guaranteed only up to $\bar{k}\sim 0.1$ as shown in the previous section.

\begin{figure}[h]
    \centering
    \includegraphics[width=1.0\linewidth]{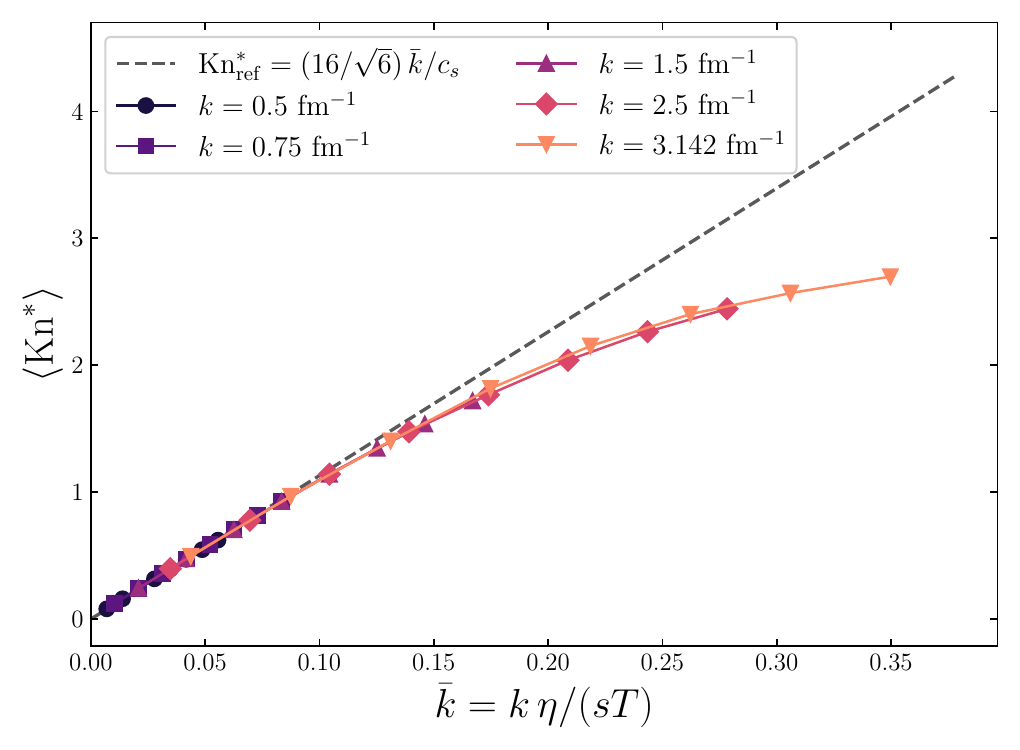}
    \caption{The $\langle \mathrm{Kn}^*\rangle$ obtained from IS hydrodynamic simulations plotted as a function of $\bar{k}$, compared to the gray dashed reference line $\Kn^*_{\rm ref} = (16/\sqrt{6})\bar{k}/c_s$. Results for different wavenumbers are shown with different filled markers. 
    The time average is taken over $0\leq t \leq 50$ fm, as in Fig.~\ref{fig:invRe_Kn_sigma_Kn_theta}.
    }
    \label{fig:Kn_star}
\end{figure}

It might be intuitively expected that the inverse Reynolds number would increase monotonically with $\bar{k}$.
This is because the inverse Reynolds number in Eq.~\eqref{eq:invRe_Knudsen} can be evaluated as, $\Rey^{-1} \sim \nu v/L \sim \nu v k  \propto \bar{k}$, where
$L$ is a typical length scale of the system and $\nu$ is a kinematic viscosity.
However, Fig.~\ref{fig:invRe_Kn_sigma_Kn_theta} shows that, for a given $\bar{k}$, $\langle \Rey^{-1} \rangle$ depends on the actual wavenumber $k$, failing to collapse onto a single function of $\bar{k}$.
This is because the velocity magnitude, $v$, which is proportional to the perturbation amplitude, enters \(\Rey^{-1}\) and \(\Kn_\theta\).
Therefore, counterintuitively, \(\langle\Rey^{-1}\rangle\) becomes smaller at larger \(\bar{k}\) or $k$ when the time average is taken over a fixed time window, because perturbations with such wavenumbers damp faster in time $t$.
As a result, both $\Rey^{-1}$ and $\Kn_\theta$ do not show the linear increase along $\bar{k}$, and also fail to converge as a single function of $\bar{k}$, when the time average is taken over a fixed time interval. See Appendix~\ref{appendix:time_average_window_dependence_of_Kntheta_and_Knstar} for a detailed discussion.
Thus, we would like to define another dimensionless number that does not directly reflect the magnitude of velocity.

The Mach number $\Maa=v/c_s$ is another dimensionless variable in hydrodynamics that is defined as the ratio of the fluid velocity to the speed of sound.
It is discussed in aerospace engineering where the compressibility of air affects the resulting force on aircraft especially at high Mach numbers.
In non-relativistic hydrodynamics, the Mach number is related to the Reynolds number and the Knudsen number through the von Kármán relation, $\Kn \propto \Maa/\Ree$ \cite{GOLSE2005159,LALLEMAND2021109713,PhysRevFluids.6.013401}. Therefore, with $\Ree \sim vL/\nu$, 
the Knudsen number is $ \Kn \sim (1/c_s) \nu/L$, where 
$\nu/c_s$ corresponds to the mean-free-path, $\lambda_{\rm mfp}$ \cite{landau1987fluid,Csernai:2011qq,Floerchinger:2011pxy}.
Note also that the kinematic viscosity is the shear viscosity $\eta$ per unit enthalpy, which becomes $\eta$ over mass density in the non-relativistic case and $\eta$ over $sT$ in the relativistic case \cite{Liao:2009gb}.
In contrast, the relativistic Reynolds number in Eq.~\eqref{eq:invRe_Knudsen} is inversely proportional to $v$, 
\footnote{
As mentioned in Ref.~\cite{Denicol:2012cn},
the relativistic Reynolds and Knudsen numbers in Eq.~\eqref{eq:invRe_Knudsen} differ from the conventional Reynolds number.
The relativistic form defines $\Ree$ as a ratio of equilibrium to non-equilibrium pressure while the conventional form defines $\Ree$ as the ratio of inertial to viscous forces in the Navier-Stokes equation.
}
since $\Ree \sim P/\sqrt{\pi^{\mu\nu} \pi_{\mu\nu}} \sim TsL/\eta v \sim L/\nu v$.
This leads to the modified von Kármán relation: $\Kn^* \sim (1/c_s^2) \Ree^{-1} /\Maa \sim (1/c_s)\nu/L$.
We denote the Knudsen number obtained from the von Kármán relation by $\Kn^*$, to
distinguish it from the Knudsen number defined in Eq.~\eqref{eq:invRe_Knudsen}.
The $\Kn^*$ can be understood as the ratio of the effective mean free path to the characteristic length scale, which is also discussed in Ref.~\cite{Liao:2009gb} as the fluidity measure.


Figure~\ref{fig:Kn_star} shows the $\langle \Kn^* \rangle = (1/c_s^2) \langle \Ree^{-1} \rangle/ \langle \Maa \rangle$ obtained from the hydrodynamic simulations as a function of $\bar{k}$
\footnote{
We define it as \(\langle \Ree^{-1}\rangle/\langle \Maa\rangle\) rather than \(\langle \Ree^{-1}\rangle \langle \Maa^{-1}\rangle\) or \( 1/(\langle \Ree\rangle \langle \Maa\rangle)\) to avoid the disproportionate influence of fluid cells with vanishingly small \(v\), whose contributions can dominate the average \(\langle \Maa^{-1}\rangle\) or \(\langle \Ree \rangle\).
},
which is also compared to the reference line $\Kn^*_{\rm ref} =(16/\sqrt{6})\bar{k}/c_s$.
The derivation of the reference $\Kn^*_{\rm ref}$ is explained in Appendix~\ref{appendix:reference_line_Knstar}. 
In $\langle \Kn^* \rangle$, the dependence on $v$ cancels. 
Consequently, the results from different parameter sets follow a common 
single function of $\bar{k}$, which is well described by the reference line especially at small $\bar{k}$.
This supports the legitimacy of the calculated $\langle \Kn^*\rangle$ regardless of the technical complications of the calibration, such as the choice of the time window or the averaging in space-time.
Contrary to $\langle \Rey^{-1} \rangle$ and $\langle \Kn_\theta \rangle$,
we find that the relation between $\langle \Kn^* \rangle$ and $\bar{k}$ 
stays robust regardless of the choice of the time-average windows (See Appendix~\ref{appendix:time_average_window_dependence_of_Kntheta_and_Knstar}).
We also observed that the relation between $\langle \Kn^* \rangle$ and $\bar{k}$ does not quantitatively change whether we start time-average from $t_{\rm min}$ or $0$ fm.

Overall, the results demonstrate that
$\langle \Kn^* \rangle$ can be a good measure for $\bar{k}$ which may assess the applicability of hydrodynamics in terms of linear responses.
From the comparisons between Figs.~\ref{fig:poles_hydro} and \ref{fig:poles_transport},
we saw that the linear responses from hydrodynamics and transport simulations start to deviate
around $\bar{k}\sim0.1$.
Therefore, $\bar{k}_{\rm max}\sim 0.1$ marks the approximate upper limit of the range in which IS hydrodynamics describes the linear response of transport simulations, corresponding to $\langle \Kn^* \rangle_{\rm max} \sim 1$ in Fig.~\ref{fig:Kn_star}. We emphasize that $\langle \Kn^* \rangle$ can be evaluated from hydrodynamic variables at the local rest frame without prior knowledge of the wavenumber of the initial perturbation.

It should be noted, however, that the presented results consider only the
linear response in sound mode around a static background, which is far from realistic simulations.
Further studies are needed to establish the practical usefulness of $\langle \Kn^* \rangle $ in realistic simulations of relativistic heavy-ion collisions.

\section{Summary and discussions}
We studied the linear response functions from numerical simulations of hydrodynamics using the PPM+HLLE algorithm and transport using JAM2 for a massless particle system.
We demonstrated that the pole locations of the linear response functions for the sound mode can be extracted by fitting the evolution of the energy density in the $k$-$t$ space.
We showed that the extracted poles with different temperatures, wavenumbers of initial perturbations, and shear viscosities from numerical simulations exhibit universality after properly scaling $t$ and $k$ to dimensionless variables.

The method of the pole extraction can also be used as a benchmark for numerical implementations of hydrodynamics and transport codes.
For hydrodynamics, numerical dissipation is known to persist depending on the discretization algorithm, and this validation can be used to help quantify numerical dissipation relative to the physical one. 
For transport as well, the comparison of the extracted poles and the analytical expectation values can be used as a benchmark test of the transport code.
For example, we showed that the test particle method brought the results significantly closer to the analytical expectation values.
Also, note that this pole extraction method can be used as a complementary check of the viscosity in a transport model, in addition to the conventional method with the Green-Kubo formula \cite{Demir:2008tr,Rose:2017bjz,Ozvenchuk:2012kh,Wesp:2011yy}.

The universality of the rescaled linear response functions in terms of $\bar{k} = k\eta/(sT)$ is useful to highlight the fundamental difference between hydrodynamics and transport regardless of the difference in kinematic viscosities.
The rescaled linear response shows consistency for $\bar{k}\lesssim 0.1$, above which hydrodynamics starts deviating from kinetic theory. 
This result indicates that $\bar{k}_{\rm max}\sim0.1$ is the upper limit of $\bar{k}$, in which the Israel-Stewart hydrodynamics and Boltzmann transport with RTA coincide in terms of linear responses.
Therefore, finding dimensionless numbers that may represent $\bar{k}$ is useful 
if they can be calculated from hydrodynamic variables without the prior knowledge about the wavenumber of the initial states.

We investigated several dimensionless numbers as a function of $\bar{k}$, and found that the Knudsen number obtained through the relativistic form of the von Kármán relation, $\langle \Kn^* \rangle = (1/c_s^2) \langle \Rey^{-1} \rangle / \langle \Maa \rangle$, 
from different wavenumbers $k$ and different values of $\eta/s$ collapsed onto a common single function of
$\bar{k}$.
This indicates that $\langle \Kn^* \rangle $ can be a good measure of $\bar{k}$.
The practical usefulness of \(\langle \Kn^* \rangle\) for simulations of relativistic heavy ion collisions comes from the fact that it can be calculated from hydrodynamic variables without prior knowledge of the wavenumbers present in the initial state. We also confirmed that the $\bar{k}$ dependence of $\langle \Kn^* \rangle $ remains robust across the time-averaging windows considered here, allowing flexibility in the choice of time interval when evaluating the applicability of hydrodynamics in simulations of relativistic heavy-ion collisions.
We found $\bar{k}_{\rm max} \sim 0.1$ corresponds to $\langle \Kn^* \rangle_{\rm max} \sim 1$. Note that
$\bar{k}_{\rm max} \sim 0.1$ corresponds to a minimum wavelength $\lambda_{\rm min} \sim 3 \ \rm fm$ when a typical temperature and viscosity of a QGP in relativistic heavy-ion collisions, $T\sim 0.3 \ \rm GeV$ and $\eta/s>1/(4\pi)$ \cite{Kovtun:2004de}, are considered.
This raises a concern on the application of hydrodynamics, especially, on small collision systems.

It should also be noted that $k\tau_R\simeq4.5$~\cite{Romatschke:2016hle,Florkowski:2017olj} sets a hard upper limit on the applicability of hydrodynamics, as the hydrodynamic sound poles disappear beyond this threshold in kinetic theory with RTA.
Using $\tau_R=5\eta/(sT)$, this threshold corresponds to $\bar{k}=k\eta/(sT)=k\tau_R/5\simeq0.9$, which is larger than $\bar{k}_{\rm max}$, as indicated by the endpoint of the analytical RTA curve in Fig.~\ref{fig:poles_transport}.
Causality conditions~\cite{Bemfica:2020xym,Plumberg:2021bme} also provide a complementary way to quantify the applicable regime of hydrodynamics and should, in principle, be considered alongside the criteria discussed here.

The logical next step would be to repeat this analysis for an expanding system to see if the criteria developed here remain meaningful. In addition, it would be insightful to investigate $\langle \Kn^* \rangle $ 
with a non-conformal lattice QCD equation of state at the initial stages of relativistic heavy-ion collisions using commonly-adopted initial conditions,
as previous comparisons between Boltzmann kinetic theory and second-order viscous hydrodynamics in non-conformal Bjorken expansion have revealed significant differences in their early time, far-from-equilibrium evolution~\cite{Kurkela:2018vqr,Kurkela:2018wud,Ambrus:2022qya,Ambrus:2022koq,Jaiswal:2021uvv}.

\begin{acknowledgments}
We thank Lipei Du, Jørgen Randrup, and Feng Yuan for variable discussions.
We also thank Harri Niemi for his advice on implementing second-order viscous hydrodynamics in (1+1) dimensions with the SHASTA algorithm, and Yasushi Nara for helpful discussions on JAM2.
We also thank Dmytro Oliinychenko 
for critical comments on the manuscript. 
This material is based upon work supported by the U.S. Department of Energy, 
Office of Science, Office of Nuclear Physics, under contract number 
DE-AC02-05CH11231.
Y.K. is supported by the Japan Science and Technology Agency (JST) as part of Adopting Sustainable Partnerships for Innovative Research Ecosystem (ASPIRE) Grant No. JPMJAP2318. 
This research used resources of the National Energy Research Scientific Computing Center (NERSC), a Department of Energy User Facility using NERSC awards NP-ERCAP0036425.
This research was supported in part by grant NSF PHY-2309135 to the Kavli Institute for Theoretical Physics (KITP).
\end{acknowledgments}

\vspace{2.5cm}

\appendix

\section{Extracted $\Ree \bar{\omega}/\bar{k}$ and $-\Imm \bar{\omega}/\bar{k}^2$ with SHASTA}
\label{appendix:SHASTA}
\begin{figure}
    \centering
    \includegraphics[width=1\linewidth]{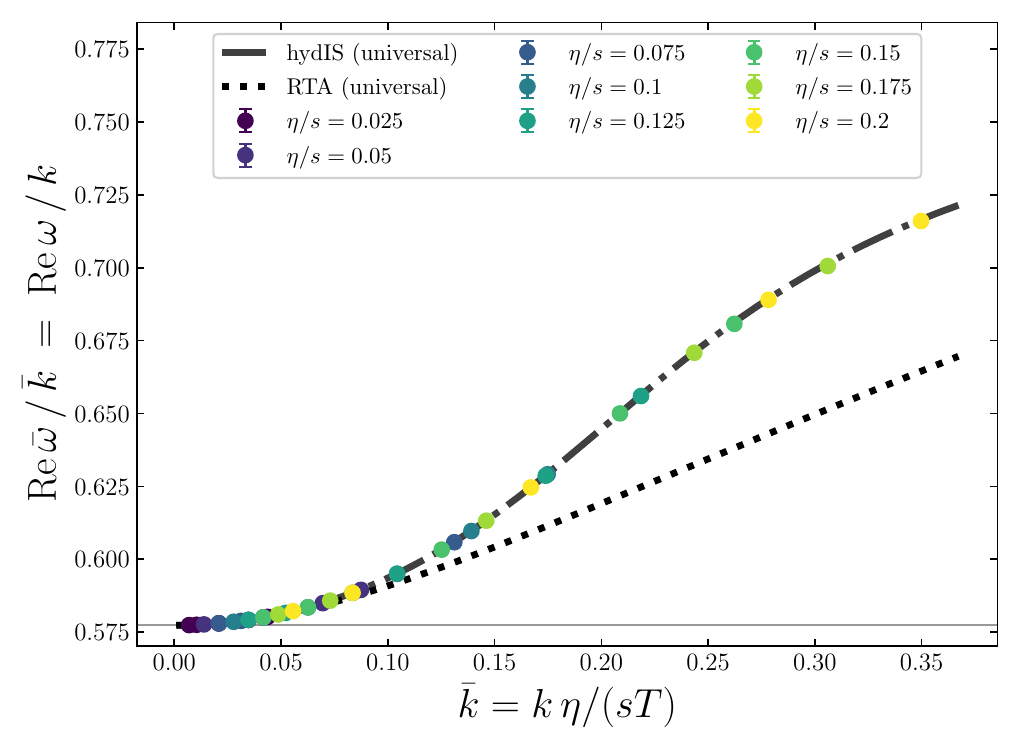}
    \includegraphics[width=1\linewidth]{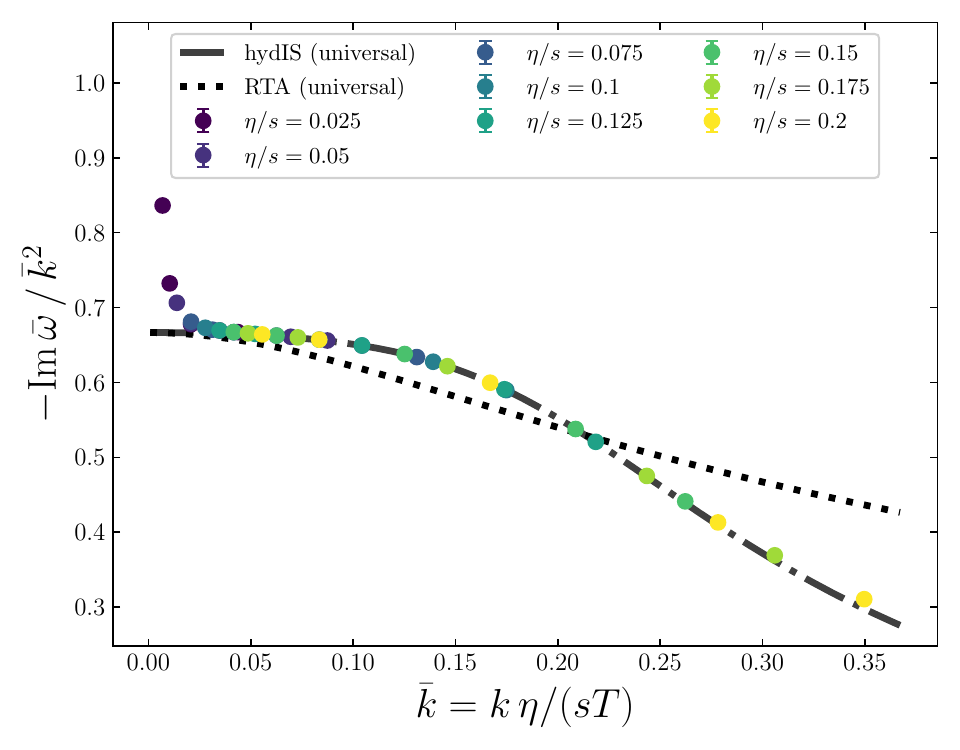}
    \caption{Same as Fig.~\ref{fig:poles_hydro}, but with results obtained using the SHASTA algorithm.}
    \label{fig:shasta_poles}
\end{figure}
To examine the dependence on the discretizations algorithm,
we performed the same analysis with hydrodynamic simulations replacing the PPM+HLLE and the predictor-corrector method with the SHASTA algorithm,
which includes both space and time discretizations.
SHASTA is a flux-corrected transport algorithm that employs a low-order diffusive transport step followed by an antidiffusion correction that removes the numerical diffusion.
The antidiffusion coefficient $A_{\rm ad}$, defined in Ref.~\cite{Molnar:2009tx}, is set to 1.0 following the discussions, for instance, in Refs.~\cite{VonGersdorff:1986tqh,Molnar:2009tx,Song:2008si}.

Figure~\ref{fig:shasta_poles} shows the results corresponding to Fig.~\ref{fig:poles_hydro}, with hydrodynamic simulations using the SHASTA algorithm. 
The results are consistent with those obtained with PPM+HLLE, clearly recovering the poles of the hydrodynamic modes of the Israel-Stewart equations.
As mentioned in Sec.~\ref{subsec:extracted_poles}, the overestimation of 
$-\Imm \bar{\omega}/\bar{k}^2$ at small $\bar{k}$ is also seen with the SHASTA algorithm, which supports the interpretation that the observed tendency has a physical origin rather than being a numerical artifact.

\begin{figure}
    \centering
    \includegraphics[width=1.0\linewidth]{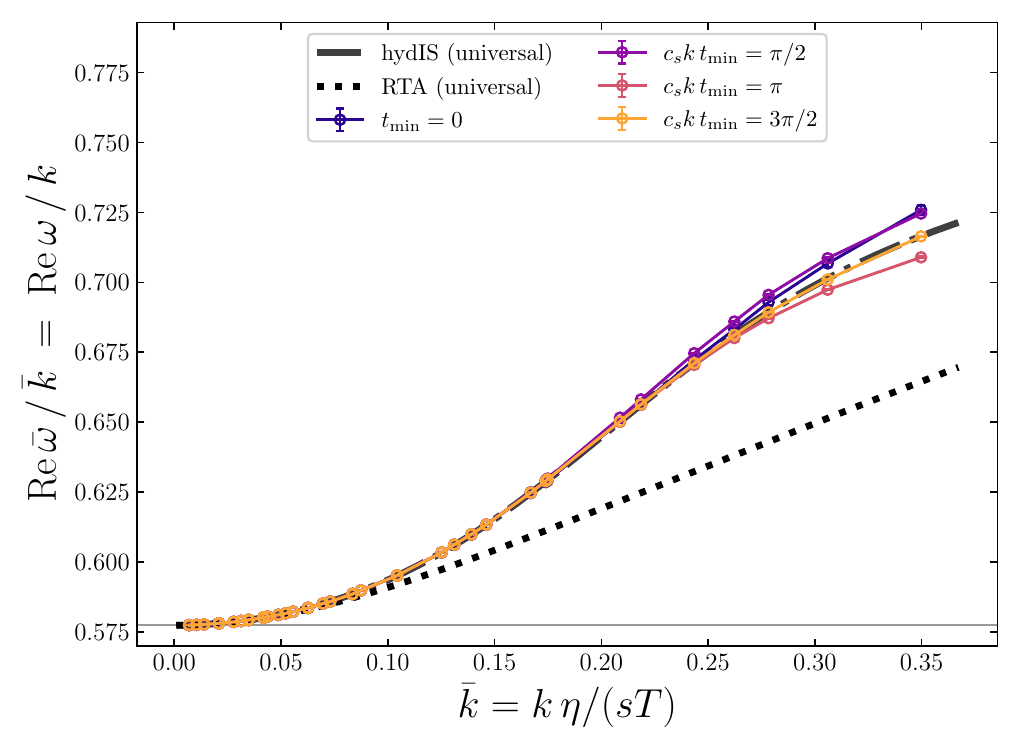}
    \includegraphics[width=1.0\linewidth]{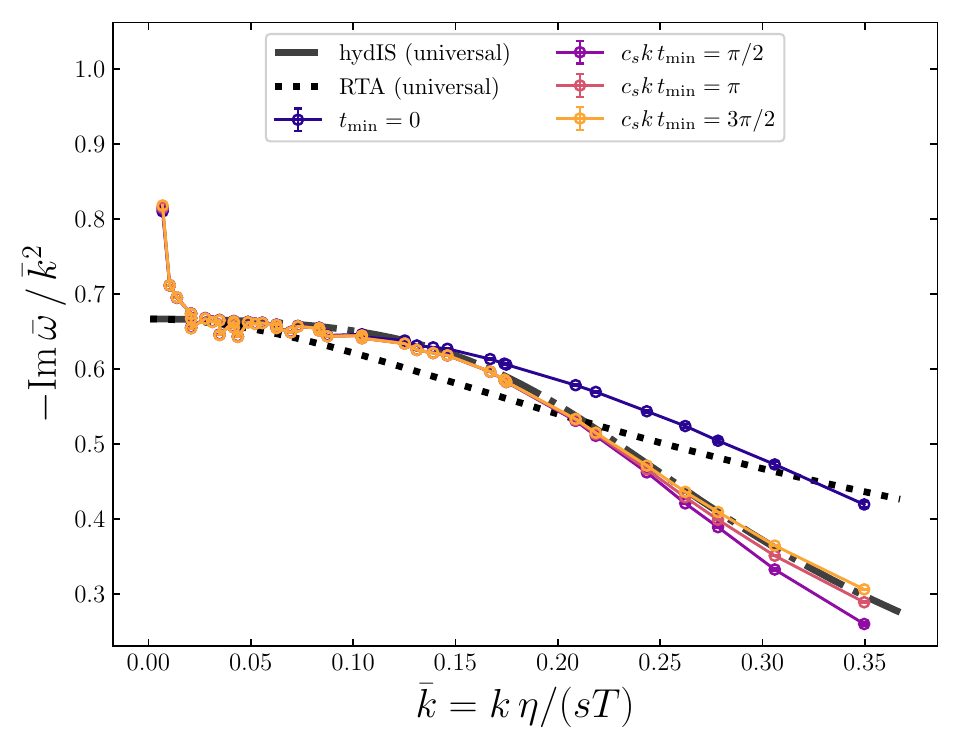}
    \caption{The $t_{\rm min}$ dependence of extracted $\Ree \bar{\omega}/\bar{k}$ and $-\Imm \bar{\omega}/\bar{k}^2$ from hydrodynamic simulations with PPM+HLLE algorithm.}
    \label{fig:pole_hydro_tmin}
\end{figure}

\begin{figure}
    \centering
    \includegraphics[width=1\linewidth]{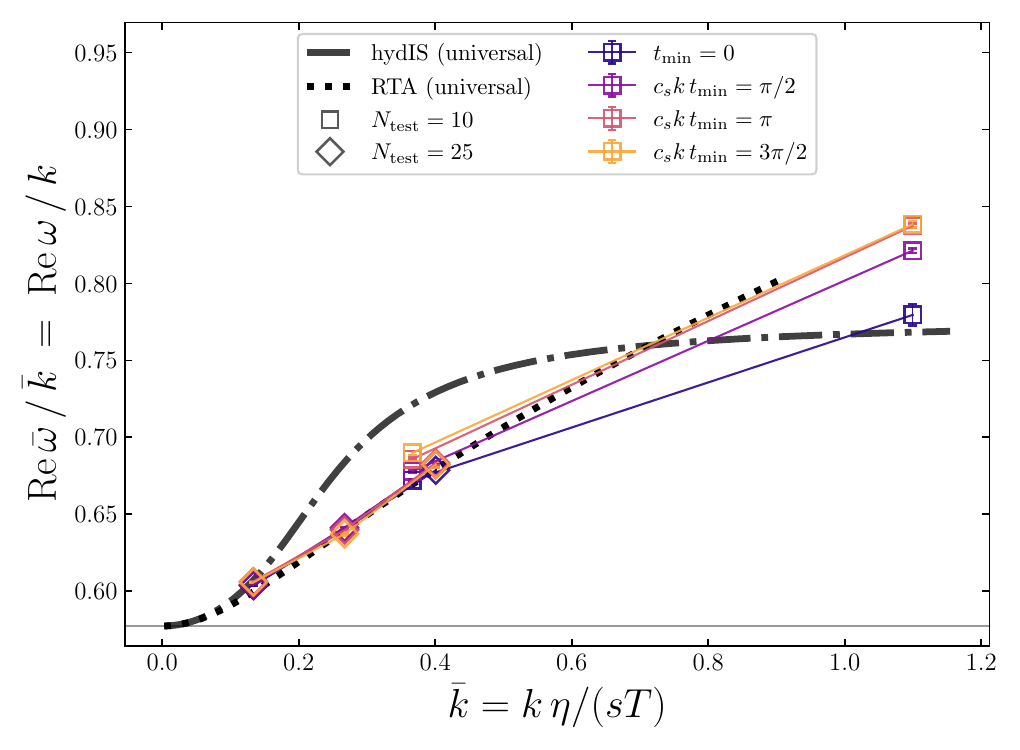}
    \includegraphics[width=1\linewidth]{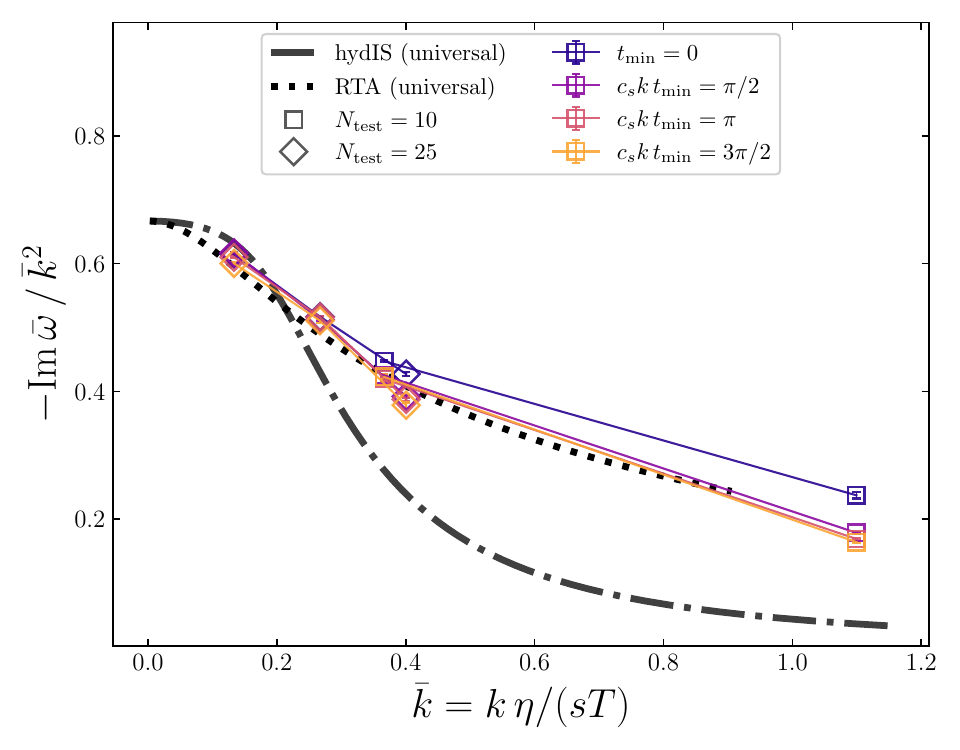}
    \caption{The $t_{\rm min}$ dependence of extracted $\Ree \bar{\omega}/\bar{k}$ and $-\Imm \bar{\omega}/\bar{k}^2$ from transport simulations. Only the results with the test particle method are shown.}
    \label{fig:pole_transport_tmin}
\end{figure}

\section{Fitting window $t_{\rm min}$ and $t_{\rm cut}$}
\label{appendix:fittings_of_deltae}

In this section, we explain the $t_{\rm min}$ dependence on the pole extraction and the method to define $t_{\rm cut}$ that avoids fits biased by statistical fluctuations in the transport results, which we mentioned in Sec.~\ref{subsec:Extraction_of_initial_value_response_function}.

Figure~\ref{fig:pole_hydro_tmin} shows the $t_{\rm min}$ dependence of the extracted $\Ree \bar{\omega}/\bar{k}$ and $-\Imm \bar{\omega}/\bar{k}^2$ from hydrodynamic simulations with the PPM+HLLE algorithm.
We choose $t_{\rm min}=0,\pi/(2c_sk),\pi/(c_sk)$, and $3\pi/(2c_sk)$, corresponding to zero, one-quarter, one-half, and three-quarters of the oscillation period $2\pi/(c_sk)$ of $\tilde{e}_k(t)$, respectively.
In Ref.~\cite{Du:2023bwi}, the hydrodynamization time $\bar{t}\sim 4\pi$ \cite{Du:2020zqg,Giacalone:2019ldn} is used as the starting time of the fitting time window.
Note that the $t_{\rm min}= 3\pi/(2c_sk)$ is much larger than the hydrodynamization time $\bar{t}\sim 4\pi$ given the parameter range.

Figure~\ref{fig:pole_transport_tmin} shows $t_{\rm min}$ dependence of the extracted $\Ree \bar{\omega}/\bar{k}$ and $-\Imm \bar{\omega}/\bar{k}^2$ from transport simulations.
We see in Fig.~\ref{fig:poles_transport} that the transport results with the test particle method recover the poles of linear response from the RTA. 
We therefore present the dependence on $t_{\rm min}$ only for the results with the test particle method to isolate the influence of early-time non-hydrodynamic modes on the extracted poles.
Both Figs.~\ref{fig:pole_hydro_tmin} and \ref{fig:pole_transport_tmin} show that the extracted poles of the linear response functions appear to converge to the analytical values for $t_{\rm min}=3\pi/(2c_sk)$, suggesting that the effect of non-hydrodynamic mode \cite{Du:2023bwi} or the numerical inaccuracies at early times is effectively eliminated in the fit.
Therefore, we use the minimum time $t_{\rm min} = 3\pi/(2c_sk)$ for the presented main results.

The minimum time $t_{\rm min}$ for the fitting time window is imposed in both hydrodynamics and transport simulation results.
The maximum time $t_{\rm cut}$ for the fitting time window is imposed only in the transport simulation results.
This is because, in the results from transport simulations, the statistical fluctuation may persist in the late-time interval although the input perturbation has fully relaxed, which may bias the extracted frequency and damping rate.

The upper limit $t_{\rm cut}$ of the time window for fitting is determined by the following algorithm.
We define the
normalized unsigned signal $S(t) \equiv |\tilde{e}(k,t)|/|\tilde{e}(k,t_0)|$
and estimate the noise floor from the last 30\% of the simulated time
interval, using its mean $\bar{S}$ and standard deviation $\sigma$. Because $S(t)$ vanishes at every zero crossing, we compare the noise with the upper envelope of $S(t)$ which is defined as,
$S_{\rm env}(t) = \max_{|t^\prime-t| \le \Delta t_{\rm node}} |S(t^\prime)|$, where
$\Delta t_{\rm node} = \pi/(c_s k)$ is the expected time interval between nodes.
The $t_{\rm cut}$ is then defined as the first time at which $S_{\rm env}$ satisfies,
\begin{align}
  S_{\rm env}(t) \le \bar{S} + n\sigma,
\end{align}
where $n=4$ is chosen in our case so that the later time window, which is dominated by statistical fluctuations, is effectively excluded.

\begin{figure}
    \centering
    \includegraphics[width=1\linewidth]{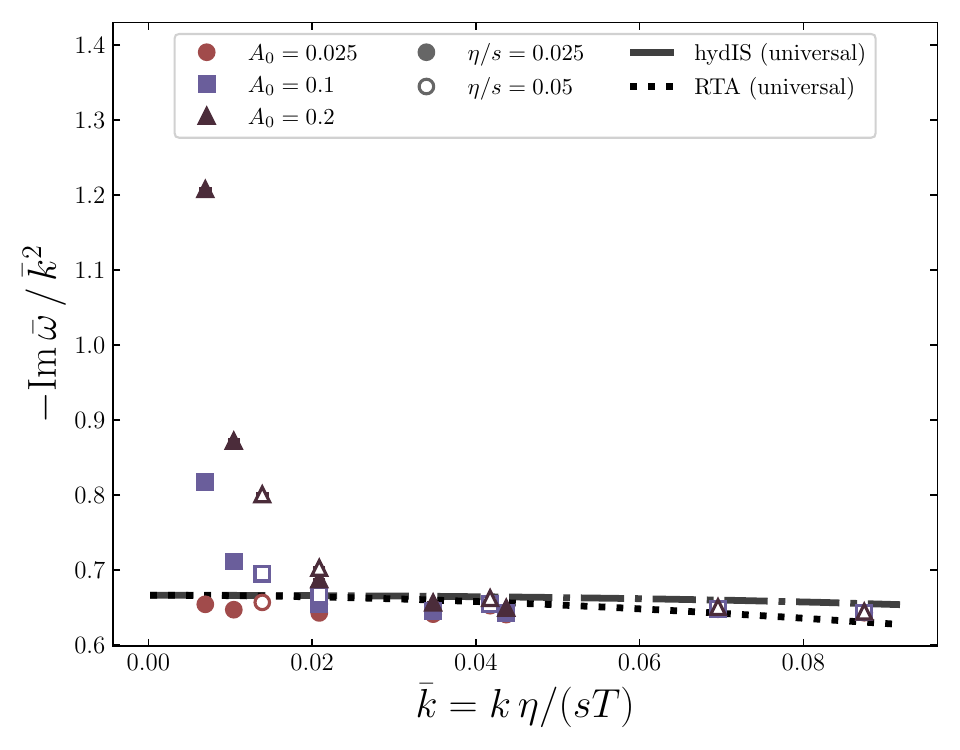}
    \caption{Dependence of $-\Imm \bar{\omega}/\bar{k}^2$, extracted from hydrodynamic simulations with PPM+HLLE algorithm, on the amplitude of the initial perturbation $A_0$.
    Results obtained from simulations with $A_0=0.025$, $0.1$, and $0.2$ using $\eta/s=0.025$ ($0.05$) are plotted in circles, squares, and triangles with filled (open) markers, respectively, which are compared to the analytical curves from the IS hydrodynamics (dash-dotted) and RTA of the Boltzmann equation (dotted). 
    }
    \label{fig:Acomparisons}
\end{figure}

\section{Amplitude dependence of the extracted $-\Imm \bar{\omega}/\bar{k}^2$ at small $\bar{k}$}
\label{appendix:amplitude_dependence_in_negative_omega}

Figure~\ref{fig:Acomparisons} shows the dependence of $-\Imm \bar{\omega}/\bar{k}^2$, which is extracted from hydrodynamic simulations with PPM+HLLE, on the amplitude of the initial perturbation $A_0$ defined in Eq.~\eqref{eq:delta_e_t0}.
Here, we show only results with $\eta/s=0.025$ and $0.05$ for clarity to investigate the cause of the deviation from the analytical curves observed in Fig.~\ref{fig:poles_hydro}.
The results in Fig.~\ref{fig:Acomparisons} show clear dependence of the extracted $-\Imm \bar{\omega}/\bar{k}^2$ on $A_0$. This demonstrates that the simulations with smaller $A_0$ bring the results closer to the analytical results, suppressing the magnitude of non-linear response.

\section{Derivation of $\Kn^*_{\rm ref} = (16/\sqrt{6})\bar{k}/c_s$}
\label{appendix:reference_line_Knstar}
The reference line $\Kn^*_{\rm ref} = (16/\sqrt{6})\bar{k}/c_s$ in Fig.~\ref{fig:Kn_star} is derived as follows.
In (1+1)D Cartesian coordinates, the shear stress tensor with the Navier-Stokes formalism shown in Eq.~\eqref{eq:NS} has the following relations:
\begin{align}
\label{eq:pipi}
    \pi^{\mu\nu} \pi_{\mu\nu} = \frac{3}{2}\pi^2,
    \quad
    \pi = \frac{4}{3} \eta \theta,
\end{align}
where $\pi$ is the only  independent variable in $\pi^{\mu\nu}$, which is defined as $\pi^{yy}=\pi^{zz}=\pi/2$, and $\theta = \partial_\mu u^\mu$ is the expansion rate.
Using $e+P = 4P = sT$, $\Kn^*$ becomes,
\begin{align}
\label{eq:Kn_star_ref}
    \Kn^* 
    & = (1/c_s^2) \ \Rey^{-1} / \Maa \nonumber \\
    & = \frac{1}{c_s v} \frac{\sqrt{\pi^{\mu\nu} \pi_{\mu \nu}}}{P} \nonumber \\
    & = \frac{1}{c_s v} \frac{16}{3sT} \sqrt{\frac{3}{2}} \eta |\theta|.
\end{align}
By replacing $|\theta|$ by $ k v$, Eq.~\eqref{eq:Kn_star_ref} becomes $\Kn^*_{\rm ref} = (16/\sqrt{6})\bar{k}/c_s$.

\begin{figure}
    \centering
    \includegraphics[width=1\linewidth]{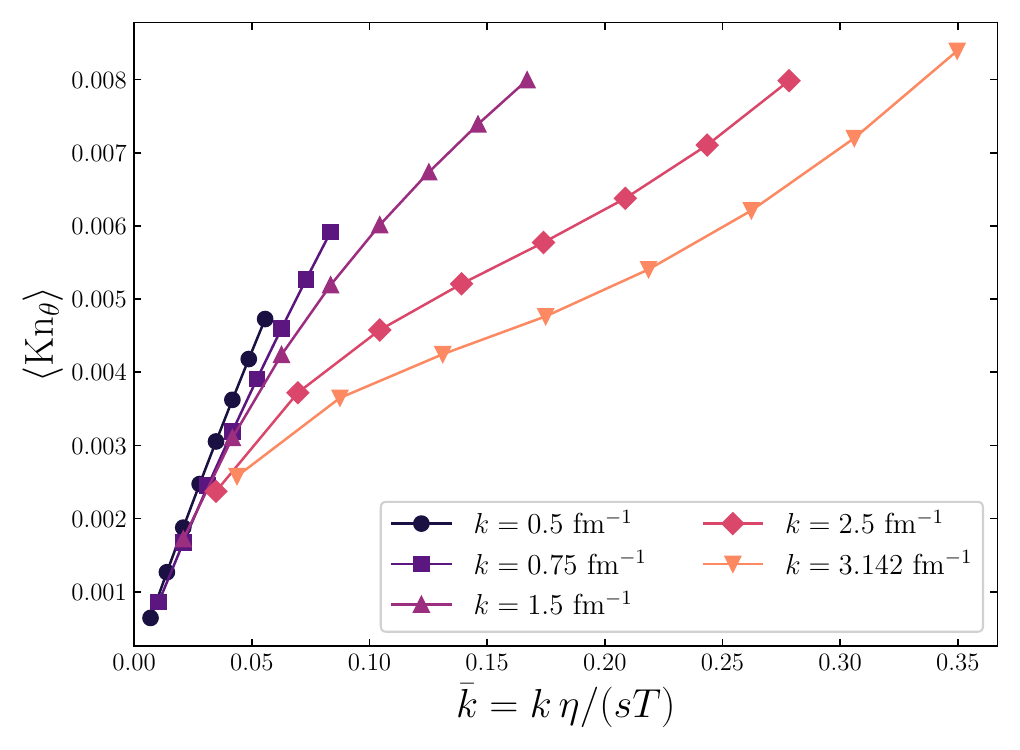}
    \includegraphics[width=1\linewidth]{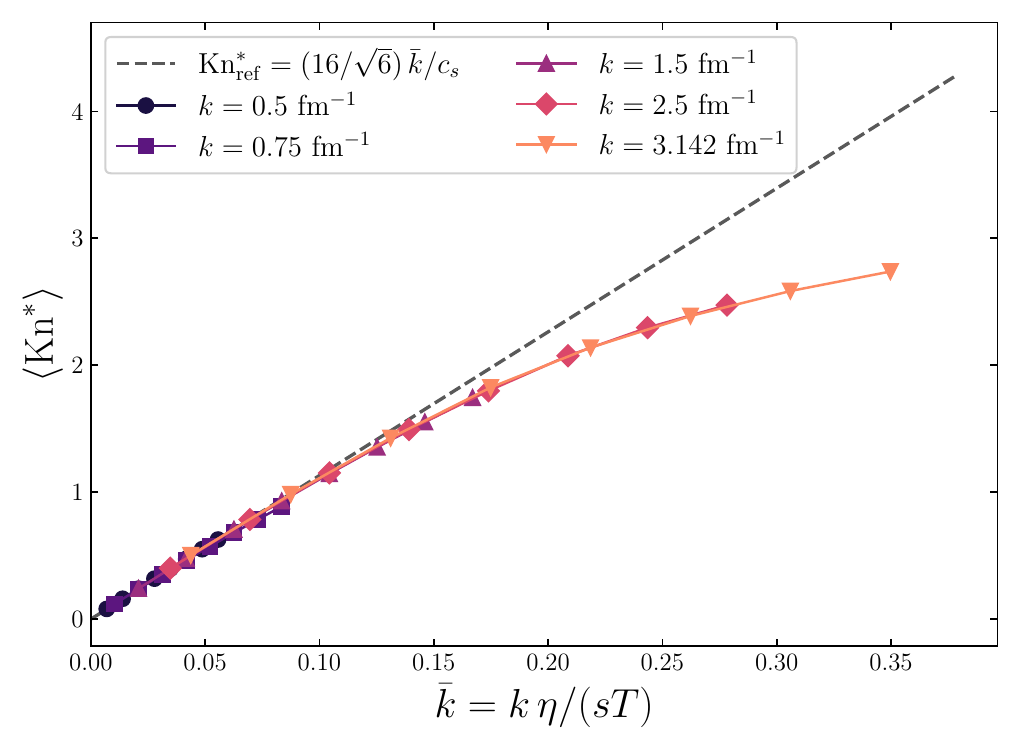}
    \caption{Space-time averaged $\Kn_\theta$ and $\Kn^*$ as functions of $\bar{k}$. The time average is taken over $0\leq t \leq 10$ fm.}
    \label{fig:Kn_Knstar_10fm}
\end{figure}
\begin{figure}
    \centering
    \includegraphics[width=1\linewidth]{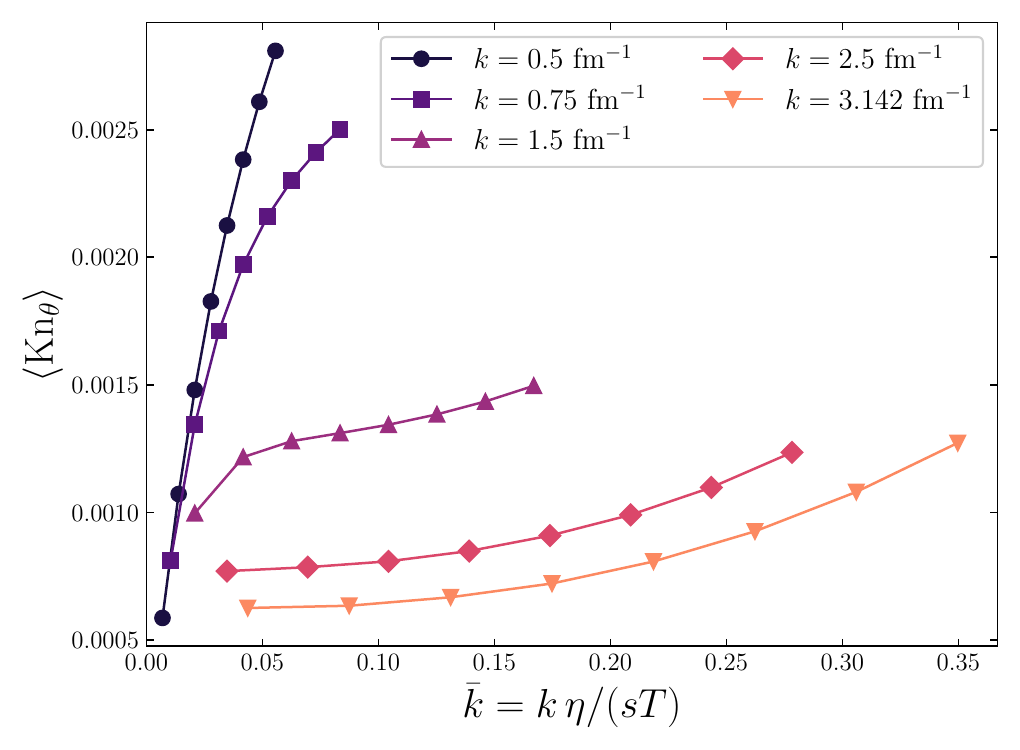}
    \includegraphics[width=1\linewidth]{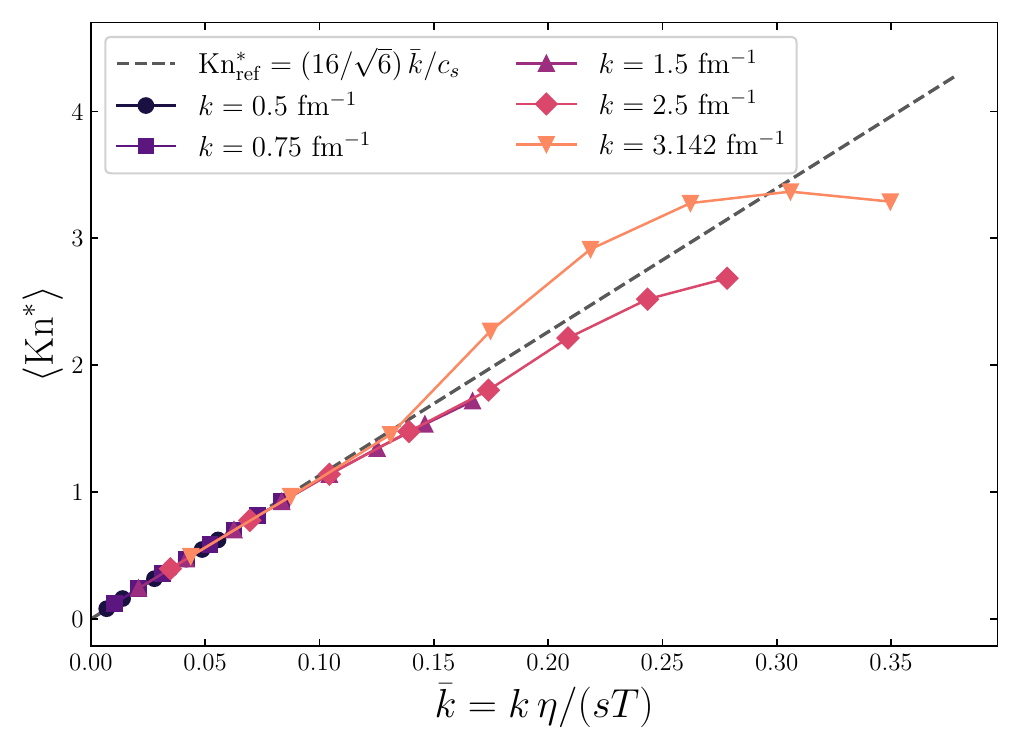}
    \caption{Same as Fig.~\ref{fig:Kn_Knstar_10fm}, but with the time-average window $0\leq t \leq 70$ fm.}
    \label{fig:Kn_Knstar_70fm}
\end{figure}

\section{Time-average window dependence on $\langle \Kn_\theta \rangle $ and $\langle \Kn^* \rangle$}
\label{appendix:time_average_window_dependence_of_Kntheta_and_Knstar}
As discussed in Sec.~\ref{subsec:DimensionLess_numbers_in_hydrodynamics}, results obtained with different parameter sets follow a common positive linear relation between $\langle \Kn^* \rangle$ and $\bar{k}$, which is well described by $\Kn^*_{\rm ref} = (16/\sqrt{6})\bar{k}/c_s$. In contrast, the other dimensionless quantities, $\langle \Rey^{-1} \rangle$ and $\langle \Kn_\theta \rangle$, defined in Eq.~\eqref{eq:invRe_Knudsen}, do not exhibit such scaling.
In this section, we show that this tendency remains robust for different time windows used for the time-average of the dimensionless variables.

Figures~\ref{fig:Kn_Knstar_10fm} and \ref{fig:Kn_Knstar_70fm} show $\langle \Kn_\theta \rangle$ and $\langle \Kn^*\rangle$ as functions of $\bar{k}$ when $\langle \Kn_\theta \rangle$ and $\langle \Kn^*\rangle$ are obtained from time-average over $0\leq t \leq 10$ fm and $0\leq t \leq 70$ fm, respectively.
Here, we only show $\langle \Kn_\theta \rangle$, not $\langle \Rey^{-1} \rangle$, since these two results are quantitatively similar as one sees from Fig.~\ref{fig:invRe_Kn_sigma_Kn_theta}.
In Figs.~\ref{fig:Kn_Knstar_10fm} and \ref{fig:Kn_Knstar_70fm}, the collapse of \(\langle \Kn^*\rangle\) onto a single curve as a function of \(\bar{k}\) remains robust across different time-averaging windows, whereas the results for \(\langle \Kn_\theta \rangle\) differ significantly over the different time windows.

In Fig.~\ref{fig:Kn_Knstar_10fm},
$\langle \Kn_\theta \rangle$ tends to increase with $\bar{k}$ for each wavenumber $k$.
This trend is more consistent across different wavenumbers than in Fig.~\ref{fig:invRe_Kn_sigma_Kn_theta}, although the results still do not collapse onto a common curve as clearly as those for $\langle \Kn^* \rangle$.
In Fig.~\ref{fig:Kn_Knstar_70fm}, the scaling of $\langle \Kn^*\rangle$ with $\bar{k}$ starts to break down at large $\bar{k}$ although overall tendency of $\langle \Kn^*\rangle$ still holds. The break down starts to appear because perturbations with large $\bar{k}$ may decay almost completely during the later part of the averaging interval, $0\leq t\leq 70$ fm. When both $\langle \Rey^{-1}\rangle$ and $\langle \Maa\rangle$ become extremely small, their ratio, $\langle \Kn^*\rangle \propto \langle \Rey^{-1}\rangle/\langle \Maa\rangle$, can become highly sensitive to small numerical fluctuations in the numerator and denominator.
Therefore, $\langle \Kn^* \rangle$ should be evaluated when the velocity amplitude remains sufficiently large compared with numerical fluctuations.
We also observe that $\langle \Kn^*\rangle$ deviates from $\Kn^*_{\rm ref}$ at large $\bar{k}$ for all the time-averaging windows shown in Figs.~\ref{fig:Kn_star}, \ref{fig:Kn_Knstar_10fm}, and \ref{fig:Kn_Knstar_70fm}. This is because the reference value $\Kn^*_{\rm ref}$ is derived using $\pi^{\mu\nu}$ in the Navier--Stokes limit, as explained in Appendix~\ref{appendix:reference_line_Knstar}. At large $\bar{k}$, however, the time-derivative term $\tau_\pi \dot{\pi}^{\mu\nu}$ in Eq.~\eqref{eq:IS} becomes significant, causing $\pi^{\mu\nu}$ to differ from the Navier--Stokes value given in Eq.~\eqref{eq:NS}.

The evaluated values of $\langle \Kn_\theta \rangle$ and $\langle \Rey^{-1} \rangle$ depend on the choice of time-averaging window, which makes them difficult to use as measures of $\bar{k}$. This dependence arises because $\Kn_\theta$ and $\Rey^{-1}$ in Eq.~\eqref{eq:invRe_Knudsen} of Sec.~\ref{subsec:DimensionLess_numbers_in_hydrodynamics} vary significantly over time within a given window.

The inverse Reynolds number and Knudsen number can be evaluated as follows:
$\Rey^{-1} \sim \nu v/L \sim \nu k v$, and $\Kn_\theta \sim \nu v/L \sim \nu kv$.
Thus, at fixed velocity amplitude \(v\), both quantities are proportional to \(\bar{k}=\nu k\).
Therefore, when $\langle \Kn_\theta \rangle$ and $\langle \Rey^{-1} \rangle$ are evaluated 
from a limited time interval, in which the velocity amplitude $v$ is similar across cases with different values of $k$ and $\eta/s$, they show a relatively clear increase proportional to $\bar{k}$ as seen in Fig.~\ref{fig:Kn_Knstar_10fm}.
Within the time window, $0\leq t\leq 10 $ fm, accounting only for the early time of the evolutions shown in Fig.~\ref{fig:Kn_Knstar_10fm}, 
the amplitudes of perturbations are similar across all the parameters considered here, because the amplitudes of perturbations $A_0$ at the initial time is always set to 0.1 as explained in Sec.~\ref{subsec:Extraction_of_initial_value_response_function}.

On the other hand, in the later time of the evolution, the velocity amplitude $v$ significantly differs among cases with different parameters because of the various damping rates. As a result, the proportionality of $\langle \Kn_\theta \rangle$ and $\langle \Rey^{-1} \rangle$ to $\bar{k}$ breaks down.
For instance, perturbations with larger $\bar{k}$ damp more rapidly over time. Moreover, even at fixed $\bar{k}$, perturbations with larger $k$ damp more rapidly in time $t$, as indicated by Eq.~\eqref{eq:dispersion_relation_NS}.
Accordingly, as the perturbation decays, $\Kn_\theta$ and $\Rey^{-1}$ decrease.
Because the time average is taken over a fixed time in $t$ across all the cases with different parameters,
the time-averaged values are biased by the damped $\Kn_\theta$ and $\Rey^{-1}$ in the later time.
This causes $\langle \Kn_\theta \rangle$ and $\langle \Rey^{-1} \rangle$ to deviate from proportionality to $\bar{k}$ and prevents the results from collapsing onto a single curve when plotted as a function of $\bar{k}$.
This is resolved in $\langle \Kn^* \rangle$ by eliminating its velocity amplitude $v$ dependence, which can be seen from the robust results over the different choices of the time average windows.


\bibliographystyle{apsrev4-2}
\bibliography{references}

\clearpage
\section*{Supplemental Material}

%




%
%

\section*{Fitting results of $\delta \tilde{e}_k(t)/ \delta \tilde{e}_k(0)$}
\label{appendix:fittings_of_deltae}
As explained in Sec.~III. B, 
we extract the poles of linear responses by fitting $\delta \tilde{e}_k(t)/ \delta \tilde{e}_k(0)$ using Eq.~(25).
The fitting minimizes the following value:
\begin{align}
S(A, \Gamma^p, \Omega^p, \phi) = \sum_{i \in \text{window}} \Big[g_i - m(t_i)\Big]^2, 
\end{align}
where
$m(t_i)=A e^{-\Gamma^p t_i}\cos(\Omega^p t_i + \phi)$, is the fitting function in Eq.~(25), and 
$g_i = \delta \tilde{e}_k(t_i)/ \delta \tilde{e}_k(0)$.
The fit results for different time windows for the hydrodynamic simulations using PPM+HLLE are listed in Tables~\ref{tab:fit_RTA_notmax_notmin} to \ref{tab:fit_RTA_notmax_tmin1.5pi}.
The corresponding results for transport simulations are listed in Tables~\ref{tab:fit_RTA_notmax_notmin} to \ref{tab:fit_LBT_tmin1.5pi}.
To evaluate the quality of the fitting, we define the relative fitting error,
\begin{align}
    \epsilon_{\rm rel} [\%] = \sqrt{\frac{\sum_i [g_i - m(t_i)]^2}{\sum_i m(t_i)^2}} \times 100,
\end{align}
and list in the tables for each dataset. 
As expected, $\epsilon_{\rm rel}$ becomes larger in larger $\bar{k}$, due to the emergence of non-hydrodynamic branch cuts.

\begin{table}[ht]
    \centering
    \begin{tabular}{ll|ccccc}
        \hline
        $k$ [fm$^{-1}$]& $\eta/s$ &  $\Omega^p$ & $\Gamma^p$ & $A_0$ & $\phi \ \rm [rad]$ & $\epsilon_{\rm rel}$ [\%] \\
        \hline
        0.5 & 0.025 & 0.288695 & 0.002818 & 1.0121 & -0.0108 & 0.4622 \\
        0.5 & 0.05 & 0.288823 & 0.004834 & 1.0021 & -0.0184 & 0.2296 \\
        0.5 & 0.075 & 0.289014 & 0.007033 & 1.0000 & -0.0260 & 0.2053 \\
        0.5 & 0.1 & 0.289278 & 0.009292 & 0.9998 & -0.0338 & 0.1623 \\
        0.5 & 0.125 & 0.289618 & 0.011572 & 1.0000 & -0.0419 & 0.1236 \\
        0.5 & 0.15 & 0.290036 & 0.013857 & 1.0004 & -0.0502 & 0.0946 \\
        0.5 & 0.175 & 0.290533 & 0.016139 & 1.0009 & -0.0587 & 0.0745 \\
        0.5 & 0.2 & 0.291110 & 0.018415 & 1.0013 & -0.0674 & 0.0613 \\
        \hline
        0.75 & 0.025 & 0.433140 & 0.005567 & 1.0025 & -0.0152 & 0.4868 \\
        0.75 & 0.05 & 0.433524 & 0.010456 & 0.9989 & -0.0260 & 0.3084 \\
        0.75 & 0.075 & 0.434161 & 0.015574 & 0.9996 & -0.0378 & 0.1668 \\
        0.75 & 0.1 & 0.435058 & 0.020720 & 1.0004 & -0.0502 & 0.1001 \\
        0.75 & 0.125 & 0.436222 & 0.025853 & 1.0011 & -0.0630 & 0.0684 \\
        0.75 & 0.15 & 0.437660 & 0.030956 & 1.0017 & -0.0764 & 0.0552 \\
        0.75 & 0.175 & 0.439380 & 0.036014 & 1.0023 & -0.0903 & 0.0577 \\
        0.75 & 0.2 & 0.441392 & 0.041013 & 1.0028 & -0.1050 & 0.0788 \\
        \hline
        1.5 & 0.025 & 0.867065 & 0.020512 & 0.9981 & -0.0261 & 0.3883 \\
        1.5 & 0.05 & 0.870139 & 0.041064 & 1.0004 & -0.0501 & 0.1021 \\
        1.5 & 0.075 & 0.875351 & 0.061529 & 1.0017 & -0.0763 & 0.0570 \\
        1.5 & 0.1 & 0.882824 & 0.081639 & 1.0029 & -0.1049 & 0.0871 \\
        1.5 & 0.125 & 0.892695 & 0.101160 & 1.0038 & -0.1364 & 0.2048 \\
        1.5 & 0.15 & 0.905059 & 0.119832 & 1.0042 & -0.1712 & 0.4455 \\
        1.5 & 0.175 & 0.919892 & 0.137392 & 1.0042 & -0.2092 & 0.8604 \\
        1.5 & 0.2 & 0.936972 & 0.153583 & 1.0038 & -0.2499 & 1.4996 \\
        \hline
        2.5 & 0.025 & 1.448185 & 0.056171 & 0.9998 & -0.0419 & 0.1480 \\
        2.5 & 0.05 & 1.462714 & 0.112945 & 1.0022 & -0.0855 & 0.0636 \\
        2.5 & 0.075 & 1.487893 & 0.167747 & 1.0041 & -0.1362 & 0.2278 \\
        2.5 & 0.1 & 1.524448 & 0.218731 & 1.0053 & -0.1955 & 0.7552 \\
        2.5 & 0.125 & 1.571479 & 0.263988 & 1.0060 & -0.2618 & 1.8657 \\
        2.5 & 0.15 & 1.625374 & 0.301805 & 1.0068 & -0.3299 & 3.7044 \\
        2.5 & 0.175 & 1.680593 & 0.330960 & 1.0074 & -0.3926 & 6.2466 \\
        2.5 & 0.2 & 1.732183 & 0.350923 & 1.0067 & -0.4450 & 9.3234 \\
        \hline
        3.1416 & 0.025 & 1.823606 & 0.088353 & 1.0005 & -0.0525 & 0.0936 \\
        3.1416 & 0.05 & 1.852943 & 0.177070 & 1.0034 & -0.1104 & 0.1207 \\
        3.1416 & 0.075 & 1.904153 & 0.260087 & 1.0057 & -0.1807 & 0.6162 \\
        3.1416 & 0.1 & 1.976491 & 0.332814 & 1.0075 & -0.2623 & 1.9590 \\
        3.1416 & 0.125 & 2.061624 & 0.390997 & 1.0099 & -0.3455 & 4.4321 \\
        3.1416 & 0.15 & 2.146211 & 0.431762 & 1.0115 & -0.4173 & 7.9261 \\
        3.1416 & 0.175 & 2.220303 & 0.454492 & 1.0099 & -0.4715 & 12.0566 \\
        3.1416 & 0.2 & 2.280518 & 0.460835 & 1.0031 & -0.5085 & 16.4332 \\
        \hline
    \end{tabular}
        \caption{Results from hydrodynamic simulations with PPM+HLLE algorithm with the fitting time window over $0\leq t \leq t_{\rm max}$.}
    \label{tab:fit_RTA_notmax_notmin}
\end{table}    

\begin{table}[ht]
    \centering
    \begin{tabular}{ll|ccccc}
        \hline
        $k$ [fm$^{-1}$]& $\eta/s$ &  $\Omega^p$ & $\Gamma^p$ & $A_0$ & $\phi \ \rm[rad]$ & $\epsilon_{\rm rel}$ [\%] \\
        \hline
        0.5 & 0.025 & 0.288703 & 0.002830 & 1.0139 & -0.0120 & 0.3764 \\
        0.5 & 0.05 & 0.288826 & 0.004838 & 1.0026 & -0.0187 & 0.2237 \\
        0.5 & 0.075 & 0.289015 & 0.007033 & 1.0000 & -0.0261 & 0.2133 \\
        0.5 & 0.1 & 0.289278 & 0.009291 & 0.9996 & -0.0338 & 0.1704 \\
        0.5 & 0.125 & 0.289618 & 0.011570 & 0.9998 & -0.0418 & 0.1302 \\
        0.5 & 0.15 & 0.290035 & 0.013854 & 1.0002 & -0.0501 & 0.0995 \\
        0.5 & 0.175 & 0.290532 & 0.016135 & 1.0006 & -0.0586 & 0.0773 \\
        0.5 & 0.2 & 0.291108 & 0.018409 & 1.0010 & -0.0673 & 0.0612 \\
        \hline
        0.75 & 0.025 & 0.433143 & 0.005570 & 1.0029 & -0.0156 & 0.4903 \\
        0.75 & 0.05 & 0.433524 & 0.010455 & 0.9987 & -0.0260 & 0.3197 \\
        0.75 & 0.075 & 0.434160 & 0.015571 & 0.9994 & -0.0378 & 0.1747 \\
        0.75 & 0.1 & 0.435057 & 0.020715 & 1.0001 & -0.0501 & 0.1051 \\
        0.75 & 0.125 & 0.436220 & 0.025846 & 1.0008 & -0.0629 & 0.0700 \\
        0.75 & 0.15 & 0.437656 & 0.030945 & 1.0013 & -0.0762 & 0.0501 \\
        0.75 & 0.175 & 0.439376 & 0.035996 & 1.0018 & -0.0902 & 0.0373 \\
        0.75 & 0.2 & 0.441388 & 0.040979 & 1.0020 & -0.1048 & 0.0288 \\
        \hline
        1.5 & 0.025 & 0.867064 & 0.020505 & 0.9978 & -0.0260 & 0.4012 \\
        1.5 & 0.05 & 0.870135 & 0.041053 & 1.0001 & -0.0501 & 0.1071 \\
        1.5 & 0.075 & 0.875344 & 0.061506 & 1.0013 & -0.0762 & 0.0502 \\
        1.5 & 0.1 & 0.882818 & 0.081565 & 1.0020 & -0.1048 & 0.0290 \\
        1.5 & 0.125 & 0.892705 & 0.100909 & 1.0012 & -0.1365 & 0.0188 \\
        1.5 & 0.15 & 0.905124 & 0.119081 & 0.9976 & -0.1716 & 0.0154 \\
        1.5 & 0.175 & 0.920089 & 0.135473 & 0.9891 & -0.2104 & 0.0365 \\
        1.5 & 0.2 & 0.937427 & 0.149346 & 0.9736 & -0.2521 & 0.1123 \\
        \hline
        2.5 & 0.025 & 1.448182 & 0.056156 & 0.9996 & -0.0418 & 0.1555 \\
        2.5 & 0.05 & 1.462704 & 0.112881 & 1.0016 & -0.0854 & 0.0417 \\
        2.5 & 0.075 & 1.487933 & 0.167276 & 1.0012 & -0.1364 & 0.0187 \\
        2.5 & 0.1 & 1.524823 & 0.216140 & 0.9926 & -0.1970 & 0.0273 \\
        2.5 & 0.125 & 1.572946 & 0.254508 & 0.9667 & -0.2666 & 0.1684 \\
        2.5 & 0.15 & 1.629001 & 0.276990 & 0.9155 & -0.3396 & 0.6897 \\
        2.5 & 0.175 & 1.686615 & 0.281353 & 0.8409 & -0.4059 & 1.9630 \\
        2.5 & 0.2 & 1.738782 & 0.270779 & 0.7563 & -0.4550 & 4.2620 \\
        \hline
        3.1416 & 0.025 & 1.823599 & 0.088326 & 1.0002 & -0.0524 & 0.0971 \\
        3.1416 & 0.05 & 1.852944 & 0.176826 & 1.0020 & -0.1104 & 0.0254 \\
        3.1416 & 0.075 & 1.904518 & 0.257659 & 0.9957 & -0.1819 & 0.0197 \\
        3.1416 & 0.1 & 1.978631 & 0.320035 & 0.9655 & -0.2679 & 0.1633 \\
        3.1416 & 0.125 & 2.067587 & 0.350855 & 0.8957 & -0.3574 & 0.9001 \\
        3.1416 & 0.15 & 2.155109 & 0.346828 & 0.7939 & -0.4297 & 2.9165 \\
        3.1416 & 0.175 & 2.226210 & 0.319940 & 0.6895 & -0.4678 & 6.5507 \\
        3.1416 & 0.2 & 2.276474 & 0.285859 & 0.6042 & -0.4697 & 11.4447 \\
        \hline
    \end{tabular}
        \caption{Results from hydrodynamic simulations with PPM+HLLE algorithm with the fitting time window over $\pi/(2c_sk)\leq t \leq t_{\rm max}$.}
    \label{tab:fit_RTA_notmax_tmin0.5pi}
\end{table}

\begin{table}[ht]
    \centering
    \begin{tabular}{ll|ccccc}
        \hline
        $k$ [fm$^{-1}$]& $\eta/s$ &  $\Omega^p$ & $\Gamma^p$ & $A_0$ & $\phi \ \rm[rad]$ & $\epsilon_{\rm rel}$ [\%] \\
        \hline
        0.5 & 0.025 & 0.288701 & 0.002836 & 1.0148 & -0.0116 & 0.3539 \\
        0.5 & 0.05 & 0.288828 & 0.004837 & 1.0024 & -0.0190 & 0.2260 \\
        0.5 & 0.075 & 0.289018 & 0.007030 & 0.9996 & -0.0265 & 0.2099 \\
        0.5 & 0.1 & 0.289282 & 0.009286 & 0.9992 & -0.0343 & 0.1638 \\
        0.5 & 0.125 & 0.289623 & 0.011565 & 0.9994 & -0.0423 & 0.1224 \\
        0.5 & 0.15 & 0.290040 & 0.013848 & 0.9998 & -0.0505 & 0.0915 \\
        0.5 & 0.175 & 0.290537 & 0.016130 & 1.0002 & -0.0589 & 0.0696 \\
        0.5 & 0.2 & 0.291113 & 0.018404 & 1.0006 & -0.0676 & 0.0541 \\
        \hline
        0.75 & 0.025 & 0.433146 & 0.005569 & 1.0027 & -0.0159 & 0.4976 \\
        0.75 & 0.05 & 0.433529 & 0.010448 & 0.9981 & -0.0265 & 0.3160 \\
        0.75 & 0.075 & 0.434167 & 0.015563 & 0.9989 & -0.0382 & 0.1675 \\
        0.75 & 0.1 & 0.435064 & 0.020707 & 0.9997 & -0.0505 & 0.0972 \\
        0.75 & 0.125 & 0.436227 & 0.025838 & 1.0004 & -0.0633 & 0.0625 \\
        0.75 & 0.15 & 0.437664 & 0.030936 & 1.0010 & -0.0766 & 0.0434 \\
        0.75 & 0.175 & 0.439383 & 0.035987 & 1.0015 & -0.0905 & 0.0318 \\
        0.75 & 0.2 & 0.441395 & 0.040971 & 1.0017 & -0.1050 & 0.0242 \\
        \hline
        1.5 & 0.025 & 0.867074 & 0.020492 & 0.9971 & -0.0266 & 0.3980 \\
        1.5 & 0.05 & 0.870150 & 0.041036 & 0.9997 & -0.0505 & 0.0993 \\
        1.5 & 0.075 & 0.875360 & 0.061488 & 1.0010 & -0.0765 & 0.0433 \\
        1.5 & 0.1 & 0.882833 & 0.081548 & 1.0017 & -0.1051 & 0.0238 \\
        1.5 & 0.125 & 0.892717 & 0.100894 & 1.0010 & -0.1366 & 0.0149 \\
        1.5 & 0.15 & 0.905122 & 0.119070 & 0.9974 & -0.1716 & 0.0099 \\
        1.5 & 0.175 & 0.920028 & 0.135472 & 0.9891 & -0.2098 & 0.0082 \\
        1.5 & 0.2 & 0.937159 & 0.149386 & 0.9740 & -0.2496 & 0.0120 \\
        \hline
        2.5 & 0.025 & 1.448204 & 0.056134 & 0.9991 & -0.0423 & 0.1483 \\
        2.5 & 0.05 & 1.462730 & 0.112856 & 1.0014 & -0.0857 & 0.0358 \\
        2.5 & 0.075 & 1.487952 & 0.167254 & 1.0010 & -0.1365 & 0.0139 \\
        2.5 & 0.1 & 1.524763 & 0.216126 & 0.9926 & -0.1966 & 0.0070 \\
        2.5 & 0.125 & 1.572237 & 0.254576 & 0.9671 & -0.2627 & 0.0211 \\
        2.5 & 0.15 & 1.625494 & 0.277575 & 0.9187 & -0.3215 & 0.1389 \\
        2.5 & 0.175 & 1.676286 & 0.283856 & 0.8534 & -0.3533 & 0.5654 \\
        2.5 & 0.2 & 1.717914 & 0.277691 & 0.7878 & -0.3470 & 1.5635 \\
        \hline
        3.1416 & 0.025 & 1.823629 & 0.088295 & 0.9998 & -0.0528 & 0.0894 \\
        3.1416 & 0.05 & 1.852974 & 0.176796 & 1.0018 & -0.1106 & 0.0208 \\
        3.1416 & 0.075 & 1.904485 & 0.257645 & 0.9957 & -0.1818 & 0.0139 \\
        3.1416 & 0.1 & 1.977747 & 0.320187 & 0.9662 & -0.2640 & 0.0211 \\
        3.1416 & 0.125 & 2.061702 & 0.352383 & 0.9020 & -0.3332 & 0.2001 \\
        3.1416 & 0.15 & 2.136578 & 0.353640 & 0.8191 & -0.3537 & 0.9736 \\
        3.1416 & 0.175 & 2.190847 & 0.337579 & 0.7498 & -0.3174 & 2.7590 \\
        3.1416 & 0.2 & 2.227232 & 0.317657 & 0.7087 & -0.2499 & 5.5151 \\
        \hline
    \end{tabular}    
    \caption{Results from hydrodynamic simulations with PPM+HLLE algorithm with the fitting time window over $\pi/(c_sk)\leq t \leq t_{\rm max}$.}
    \label{tab:fit_RTA_notmax_tminpi}
\end{table}

\begin{table}[ht]
    \centering
    \begin{tabular}{ll|ccccc}
        \hline
        $k$ [fm$^{-1}$]& $\eta/s$ &  $\Omega^p$ & $\Gamma^p$ & $A_0$ & $\phi \ \rm[rad]$ & $\epsilon_{\rm rel}$ [\%] \\
        \hline
        0.5 & 0.025 & 0.288707 & 0.002845 & 1.0163 & -0.0126 & 0.2885 \\
        0.5 & 0.05 & 0.288829 & 0.004838 & 1.0025 & -0.0192 & 0.2315 \\
        0.5 & 0.075 & 0.289017 & 0.007028 & 0.9993 & -0.0264 & 0.2158 \\
        0.5 & 0.1 & 0.289280 & 0.009283 & 0.9988 & -0.0340 & 0.1663 \\
        0.5 & 0.125 & 0.289620 & 0.011560 & 0.9990 & -0.0420 & 0.1224 \\
        0.5 & 0.15 & 0.290037 & 0.013844 & 0.9994 & -0.0502 & 0.0899 \\
        0.5 & 0.175 & 0.290533 & 0.016125 & 0.9998 & -0.0587 & 0.0670 \\
        0.5 & 0.2 & 0.291110 & 0.018399 & 1.0003 & -0.0674 & 0.0508 \\
        \hline
        0.75 & 0.025 & 0.433147 & 0.005569 & 1.0028 & -0.0160 & 0.5077 \\
        0.75 & 0.05 & 0.433527 & 0.010444 & 0.9977 & -0.0263 & 0.3210 \\
        0.75 & 0.075 & 0.434163 & 0.015557 & 0.9984 & -0.0380 & 0.1679 \\
        0.75 & 0.1 & 0.435059 & 0.020700 & 0.9993 & -0.0502 & 0.0954 \\
        0.75 & 0.125 & 0.436222 & 0.025830 & 1.0000 & -0.0630 & 0.0595 \\
        0.75 & 0.15 & 0.437659 & 0.030928 & 1.0007 & -0.0763 & 0.0398 \\
        0.75 & 0.175 & 0.439377 & 0.035979 & 1.0012 & -0.0902 & 0.0281 \\
        0.75 & 0.2 & 0.441388 & 0.040964 & 1.0015 & -0.1048 & 0.0205 \\
        \hline
        1.5 & 0.025 & 0.867068 & 0.020482 & 0.9966 & -0.0262 & 0.4025 \\
        1.5 & 0.05 & 0.870141 & 0.041023 & 0.9993 & -0.0502 & 0.0975 \\
        1.5 & 0.075 & 0.875349 & 0.061474 & 1.0007 & -0.0763 & 0.0401 \\
        1.5 & 0.1 & 0.882821 & 0.081534 & 1.0015 & -0.1048 & 0.0207 \\
        1.5 & 0.125 & 0.892704 & 0.100882 & 1.0008 & -0.1364 & 0.0124 \\
        1.5 & 0.15 & 0.905110 & 0.119062 & 0.9973 & -0.1715 & 0.0076 \\
        1.5 & 0.175 & 0.920015 & 0.135468 & 0.9891 & -0.2096 & 0.0056 \\
        1.5 & 0.2 & 0.937152 & 0.149404 & 0.9742 & -0.2495 & 0.0047 \\
        \hline
        2.5 & 0.025 & 1.448191 & 0.056109 & 0.9986 & -0.0420 & 0.1476 \\
        2.5 & 0.05 & 1.462710 & 0.112827 & 1.0010 & -0.0854 & 0.0319 \\
        2.5 & 0.075 & 1.487934 & 0.167234 & 1.0009 & -0.1364 & 0.0124 \\
        2.5 & 0.1 & 1.524748 & 0.216121 & 0.9925 & -0.1965 & 0.0063 \\
        2.5 & 0.125 & 1.572239 & 0.254665 & 0.9677 & -0.2627 & 0.0060 \\
        2.5 & 0.15 & 1.625717 & 0.278320 & 0.9229 & -0.3230 & 0.0229 \\
        2.5 & 0.175 & 1.677787 & 0.286870 & 0.8691 & -0.3631 & 0.1268 \\
        2.5 & 0.2 & 1.723209 & 0.284908 & 0.8232 & -0.3816 & 0.4535 \\
        \hline
        3.1416 & 0.025 & 1.823613 & 0.088265 & 0.9994 & -0.0526 & 0.0874 \\
        3.1416 & 0.05 & 1.852954 & 0.176768 & 1.0015 & -0.1105 & 0.0181 \\
        3.1416 & 0.075 & 1.904453 & 0.257619 & 0.9955 & -0.1816 & 0.0111 \\
        3.1416 & 0.1 & 1.977725 & 0.320264 & 0.9666 & -0.2639 & 0.0075 \\
        3.1416 & 0.125 & 2.062129 & 0.353685 & 0.9077 & -0.3355 & 0.0426 \\
        3.1416 & 0.15 & 2.140009 & 0.359512 & 0.8428 & -0.3712 & 0.3186 \\
        3.1416 & 0.175 & 2.202323 & 0.350602 & 0.7998 & -0.3769 & 1.0974 \\
        3.1416 & 0.2 & 2.250779 & 0.336533 & 0.7805 & -0.3753 & 2.4841 \\
        \hline
    \end{tabular}
        \caption{Results from hydrodynamic simulations with PPM+HLLE algorithm with the fitting time window over $3\pi/(2c_sk)\leq t \leq t_{\rm max}$.}
    \label{tab:fit_RTA_notmax_tmin1.5pi}
\end{table}

\begin{table}[ht]
    \centering
    \begin{tabular}{lll|ccccc}
        \hline
        $k$ [fm$^{-1}$]& $T$ \rm[GeV] & $N_{\rm test}$ & $\Omega^p$ & $\Gamma^p$ & $A$ & $\phi$ \ \rm[rad]& $\epsilon_{\rm rel}$ [\%] \\
        \hline
        1.047 & 0.25 & 1 & 0.709529 & 0.168393 & 1.0228 & -0.4464 & 4.6033 \\
        1.047 & 0.25 & 10 & 0.703888 & 0.171677 & 1.0335 & -0.4372 & 3.4724 \\
        1.047 & 0.35 & 1 & 0.634304 & 0.094583 & 1.0109 & -0.1992 & 1.4953 \\
        1.047 & 0.35 & 25 & 0.632048 & 0.086010 & 1.0036 & -0.1749 & 1.9036 \\
        1.047 & 0.45 & 1 & 0.614608 & 0.060508 & 1.0049 & -0.1082 & 1.5053 \\
        \hline
        2.094 & 0.25 & 1 & 1.561688 & 0.504999 & 1.1056 & -0.6523 & 6.2493 \\
        2.094 & 0.35 & 1 & 1.350243 & 0.310881 & 1.0282 & -0.3584 & 2.5590 \\
        2.094 & 0.35 & 25 & 1.340955 & 0.289115 & 1.0305 & -0.3313 & 1.4380 \\
        2.094 & 0.45 & 1 & 1.267910 & 0.220266 & 1.0177 & -0.2266 & 1.4534 \\
        \hline
        3.142 & 0.25 & 1 & 2.447854 & 0.837723 & 1.1502 & -0.7645 & 9.1606 \\
        3.142 & 0.25 & 10 & 2.448855 & 0.819929 & 1.1393 & -0.7470 & 8.4109 \\
        3.142 & 0.35 & 1 & 2.178074 & 0.614576 & 1.0642 & -0.5117 & 3.5965 \\
        3.142 & 0.35 & 25 & 2.131767 & 0.537590 & 1.0455 & -0.4646 & 3.4062 \\
        3.142 & 0.45 & 1 & 1.979066 & 0.446825 & 1.0386 & -0.3331 & 3.4571 \\
        \hline
    \end{tabular}
    \caption{Results from transport simulations with the fitting time window over $0\leq t \leq t_{\rm cut}$.}
    \label{tab:fit_LBT_tmin0}
\end{table}

\begin{table}[ht]
    \centering
    \begin{tabular}{lll|ccccc}
        \hline
        $k$ [fm$^{-1}$] & $T$ \rm[GeV] & $N_{\rm test}$ & $\Omega^p$ & $\Gamma^p$ & $A$ & $\phi$ \ \rm [rad]& $\epsilon_{\rm rel}$ [\%] \\
        \hline
        1.047 & 0.25 & 1 & 0.714628 & 0.152883 & 0.9148 & -0.4792 & 4.4466 \\
        1.047 & 0.25 & 10 & 0.709801 & 0.163154 & 0.9765 & -0.4767 & 3.7396 \\
        1.047 & 0.35 & 1 & 0.634855 & 0.094958 & 1.0160 & -0.2062 & 1.9090 \\
        1.047 & 0.35 & 25 & 0.632662 & 0.086265 & 1.0075 & -0.1831 & 2.3502 \\
        1.047 & 0.45 & 1 & 0.614964 & 0.060849 & 1.0111 & -0.1147 & 1.6600 \\
        \hline
        2.094 & 0.25 & 1 & 1.618177 & 0.420402 & 0.8784 & -0.7922 & 2.7421 \\
        2.094 & 0.35 & 1 & 1.352915 & 0.299841 & 0.9857 & -0.3671 & 3.3538 \\
        2.094 & 0.35 & 25 & 1.343268 & 0.286660 & 1.0215 & -0.3406 & 1.8550 \\
        2.094 & 0.45 & 1 & 1.269255 & 0.221021 & 1.0226 & -0.2339 & 1.9257 \\
        \hline
        3.142 & 0.25 & 1 & 2.594313 & 0.610991 & 0.7761 & -0.9987 & 6.5642 \\
        3.142 & 0.25 & 10 & 2.580153 & 0.617110 & 0.7990 & -0.9575 & 3.6156 \\
        3.142 & 0.35 & 1 & 2.201720 & 0.558970 & 0.9469 & -0.5554 & 2.0759 \\
        3.142 & 0.35 & 25 & 2.143148 & 0.493502 & 0.9429 & -0.4864 & 2.3154 \\
        3.142 & 0.45 & 1 & 1.984539 & 0.445190 & 1.0366 & -0.3484 & 5.4250 \\
        \hline
    \end{tabular}
    \caption{Results from transport simulations with the fitting time window over $\pi/(2c_sk)\leq t \leq t_{\rm cut}$.}
    \label{tab:fit_LBT_tmin0.5pi}
\end{table}

\begin{table}[ht]
    \centering
    \begin{tabular}{lll|ccccc}
        \hline
        $k$ [fm$^{-1}$]& $T$ \rm[GeV] & $N_{\rm test}$ & $\Omega^p$ & $\Gamma^p$ & $A$ & $\phi$ \ \rm[rad]& $\epsilon_{\rm rel}$ [\%] \\
        \hline
        1.047 & 0.25 & 1 & 0.718422 & 0.150165 & 0.8864 & -0.5210 & 5.9124 \\
        1.047 & 0.25 & 10 & 0.718193 & 0.160950 & 0.9526 & -0.5624 & 4.4214 \\
        1.047 & 0.35 & 1 & 0.636623 & 0.094358 & 1.0067 & -0.2315 & 1.9797 \\
        1.047 & 0.35 & 25 & 0.634820 & 0.085304 & 0.9919 & -0.2159 & 2.2814 \\
        1.047 & 0.45 & 1 & 0.615857 & 0.060463 & 1.0031 & -0.1325 & 1.4908 \\
        \hline
        2.094 & 0.25 & 1 & 1.634299 & 0.414847 & 0.8542 & -0.8639 & 3.7885 \\
        2.094 & 0.35 & 1 & 1.352800 & 0.299264 & 0.9826 & -0.3668 & 4.4300 \\
        2.094 & 0.35 & 25 & 1.339749 & 0.288740 & 1.0341 & -0.3207 & 2.3008 \\
        2.094 & 0.45 & 1 & 1.272399 & 0.220935 & 1.0218 & -0.2535 & 2.1778 \\
        \hline
        3.142 & 0.25 & 1 & 2.657896 & 0.567034 & 0.6704 & -1.1950 & 8.5046 \\
        3.142 & 0.25 & 10 & 2.630978 & 0.583767 & 0.7171 & -1.1121 & 2.9869 \\
        3.142 & 0.35 & 1 & 2.224183 & 0.548872 & 0.9136 & -0.6281 & 2.2710 \\
        3.142 & 0.35 & 25 & 2.146628 & 0.487751 & 0.9236 & -0.4997 & 3.2253 \\
        3.142 & 0.45 & 1 & 2.000007 & 0.440856 & 1.0182 & -0.4055 & 7.0553 \\
        \hline
    \end{tabular}
    \caption{Results from transport simulations with the fitting time window over $\pi/(c_sk)\leq t \leq t_{\rm cut}$.}
    \label{tab:fit_LBT_tmin1pi}
\end{table}

\begin{table}[ht]
    \centering
    \begin{tabular}{lll|ccccc}
        \hline
        $k$ [fm$^{-1}$]& $T$ \rm[GeV] & $N_{\rm test}$ & $\Omega^p$ & $\Gamma^p$ & $A$ & $\phi$ \ \rm[rad]& $\epsilon_{\rm rel}$ [\%] \\
        \hline
        1.047 & 0.25 & 1 & 0.705440 & 0.136450 & 0.7407 & -0.3420 & 6.2881 \\
        1.047 & 0.25 & 10 & 0.722418 & 0.162219 & 0.9712 & -0.6209 & 6.3497 \\
        1.047 & 0.35 & 1 & 0.636822 & 0.094122 & 1.0030 & -0.2350 & 2.5942 \\
        1.047 & 0.35 & 25 & 0.634392 & 0.083964 & 0.9688 & -0.2080 & 2.7276 \\
        1.047 & 0.45 & 1 & 0.616173 & 0.060591 & 1.0062 & -0.1399 & 1.6850 \\
        \hline
        2.094 & 0.25 & 1 & 1.634039 & 0.411461 & 0.8392 & -0.8619 & 5.7291 \\
        2.094 & 0.35 & 1 & 1.349094 & 0.294119 & 0.9494 & -0.3408 & 6.6741 \\
        2.094 & 0.35 & 25 & 1.334551 & 0.285674 & 1.0115 & -0.2840 & 3.1898 \\
        2.094 & 0.45 & 1 & 1.274975 & 0.225225 & 1.0567 & -0.2746 & 2.7822 \\
        \hline
        3.142 & 0.25 & 1 & 2.585435 & 0.525039 & 0.5759 & -0.9041 & 9.8116 \\
        3.142 & 0.25 & 10 & 2.633876 & 0.566983 & 0.6738 & -1.1261 & 3.4590 \\
        3.142 & 0.35 & 1 & 2.213833 & 0.530326 & 0.8486 & -0.5835 & 2.7497 \\
        3.142 & 0.35 & 25 & 2.143160 & 0.476341 & 0.8810 & -0.4843 & 4.4050 \\
        3.142 & 0.45 & 1 & 1.999636 & 0.431522 & 0.9782 & -0.4031 & 10.6167 \\
        \hline
    \end{tabular}
    \caption{Results from transport simulations with the fitting time window over $3\pi/(2c_sk)\leq t \leq t_{\rm cut}$.}
    \label{tab:fit_LBT_tmin1.5pi}
\end{table}



\end{document}